\documentclass[12pt,oneside]{autart}
\usepackage[latin9]{inputenc}
\usepackage{amsmath}
\usepackage{amssymb}

\makeatletter
\newtheorem{notation}[thm]{Notation}
\newtheorem{assumption}[thm]{Assumption}

\@ifundefined{date}{}{\date{}}
\DeclareRobustCommand*\cal{\@fontswitch\relax\mathcal}

\makeatother

\begin{document}
\sloppy\begin{frontmatter}

\title{Stability of Kalman Filtering with a Random Measurement Equation: Application to Sensor Scheduling with Intermittent Observations}

\author[gua,arg]{Dami\'{a}n Marelli}\ead{damian.marelli@newcastle.edu.au}, 
\author[dal]{Tianju Sui}\ead{suitj@mail.dlut.edu.cn}, 
\author[abb]{Eduardo Rohr}\ead{eduardo.rohr@ch.abb.com}, 
\author[new,gua]{Minyue Fu}\ead{minyue.fu@newcastle.edu.au}

\address[gua]{School of Automation, Guandong University of Technology, Guangzhou, China.} 
\address[arg]{French-Argentinean International Center for Information and Systems Sciences, National Scientific and Technical Research Council, Rosario 2000, Argentina.}
\address[dal]{School of Control Science and Engineering, Dalian University of Technology, Dalian, China.}
\address[abb]{ABB Corporate Research Center, D\"attwil, 5405, Switzerland.} 
\address[new]{School of Electrical Engineering and Computer Science, University of Newcastle, University Drive, Callaghan, NSW 2308, Australia.}

\thanks{This work was supported by National Natural Science Foundation of China(61803068) and  China Postdoctoral Science Foundation (2017M621134).}

\begin{abstract}
Studying the stability of the Kalman filter whose measurements are randomly lost has been an active research topic for over a decade. In this paper we extend the existing results to a far more general setting in which the measurement equation, i.e., the measurement matrix and the measurement error covariance, are random. Our result also generalizes existing ones in the sense that it does not require the system matrix to be diagonalizable. For this general setting, we state a necessary and a sufficient condition for stability, and address their numerical computation. An important application of our generalization is a networking setting with multiple sensors which transmit their measurements to the estimator using a sensor scheduling protocol over a lossy network. We demonstrate how our result is used for assessing the stability of a Kalman filter in this multi-sensor setting.
\end{abstract}


\end{frontmatter}

\section{Introduction}

With the fast development of communications technologies, distributed
control and monitoring are becoming increasingly popular. Packet losses
resulting from communication links cause that the estimation accuracy
of a Kalman filter deteriorates. Motivated by this, the stability
condition of a Kalman filter when measurements are intermittently
available has attracted a great attention in the recent years. In~\cite{sinopoli2004kfi},
the authors established the mathematical foundations for the estimation
stability with measurement losses and pointed out that the covariance
of the estimation error may not reach a steady state. Inspired by
this, several authors have studied different aspects of the problem,
using different assumptions on network models and protocols.

When a Kalman filter is subject to randomly intermittent observations
(KFIO), its error covariance (EC) matrix becomes also random. Its
asymptotic expected value, denoted by AEEC (for asymptotic expected
error covariance), is typically used as a performance measure for
designing the components of the system, the communication channel,
and the estimator. There exists rich literature dedicated to finding
the stability conditions of the KFIO~\cite{sinopoli2004kfi,xie2012scb,huang2007skf,you2011mss,xie2008srr,mo2012kfi,plarre2009kfd,mo2008ccv,xie2007pcs,rohr2010spe,rohr2010ecd,liu2004kfp,dana2007ecn,schenato2007fce,schenato2008oen,rohr2011kfc,park2011ikf,quevedo2012ses}.
Some authors adopt the stability criterion used in~\cite{sinopoli2004kfi},
namely, a KFIO is said to be stable if its AEEC is finite~\cite{sinopoli2004kfi},
and unstable otherwise. Other authors adopt the concept of peak error
covariance introduced in~\cite{huang2007skf}. More recently, the
equivalence between the two notions of stability has been studied
in~\cite{xie2012scb,you2011mss}.

In spite of the fact that there are many papers studying stability
conditions of the KFIO, a necessary and sufficient condition for a
general system is still not available. Most answers are only partial,
in the sense that they depend on a particular structure of the system,
or offer only a sufficient condition which is not necessary. In these
papers, there are mainly two kinds of packet loss models: The first
one considers the dropouts as a sequence of independent and identically
distributed (i.i.d.) binary random variables. The second one is known
as the Gilbert-Elliott model~\cite{gilbert1960cbn,elliott1963eer},
and models the dropouts using a first-order Markov process. A generalization
of these two models is the stationary high order Markov model, also
known as finite state Markov channel (FSMC)~\cite{sadeghi2008fsm}.
It permits modeling more complex channels with memory and fading.
In the context of KFIO, this network model has been considered in~\cite{censi2011kfi},
although not for the purposes of assessing stability.

For the i.i.d. packet loss model, the authors of~\cite{sinopoli2004kfi}
showed that there exists a critical value, such that the AEEC is bounded
if the packet arrival probability is strictly greater than this value,
and unbounded if the packet arrival probability falls below the critical
value. They also provided lower and upper bounds on the critical measurement
packet arrival probability. These bounds are only tight for systems
whose observation matrix $\mathbf{C}$ is invertible, leading to a
necessary a sufficient condition for stability of this kind of systems.
This assumption was relaxed in~\cite{plarre2009kfd}, so that only
the part of the matrix $\mathbf{C}$ corresponding to the observable
subspace is requiring to be invertible. The assumption was further
relaxed by Mo et al. in~\cite{mo2008ccv}, where they studied the
case where the unstable eigenvalues of $\mathbf{A}$ have different
magnitudes.

For the Gilbert-Elliott network model, the first work studying the
stability of a KFIO is~\cite{huang2007skf}. In that work, a sufficient
condition for stability was derived, considering the peak covariance
criterion. For a scalar system, the authors showed that this sufficient
condition is also necessary. In~\cite{xie2008srr}, a new sufficient
condition for the stability of the peak covariance was established.
In the particular case where the observation matrix $\mathbf{C}$
has full column rank (FCR), the sufficient condition is also necessary.
In the case of second order systems, You et al.~\cite{you2011mss}
gave a necessary and sufficient condition for stability. In~\cite{mo2012kfi},
the authors derived a necessary and sufficient stability condition
for a kind of systems which they call \emph{non-degenerate}. This
result generalizes most necessary and sufficient stability conditions
of systems using the Gilbert-Elliott network model.

For the general FSMC model, to the best of the authors knowledge,
the only available work is~\cite{rohr2014kalman}. The authors provided
a necessary and a sufficient condition for the stability of the KFIO,
which is valid under the assumption that the state transition matrix
$\mathbf{A}$ is diagonalizable. This is the most general result known
so far, since, as the authors show, every other available result follows
as a particular case.

The goal of this paper is to generalize the result in~\cite{rohr2014kalman},
so that the resulting necessary and sufficient condition for stability
can be applied to distributed Kalman filtering problems under a much
more general setting. An important example is the state estimation
on discrete-time systems, whose measurements are acquired by multiple
sensors, and transmitted to the estimator using sensor scheduling
over a lossy network. More precisely, in some wireless networked applications,
only limited energy and bandwidth is available for data collection
and transmission. Consequently, it is not feasible that all sensors
transmit their measurements at every sampling time. Then, a method
is required to decide which sensor sends its measurement to the estimator
at each time. This decision-making process is referred to as \emph{sensor
scheduling}.

Sensor scheduling has been an active research problem for several
years. For example, Walsh and Ye~\cite{walsh2002snc} have studied
the stability for the close-loop control problem with sensor scheduling.
Also, Gupta \emph{et al~}\cite{gupta2006sss} proposed a stochastic
scheduling strategy for the networked state estimation problem, and
derived the optimal probability distribution for selecting sensors
at each sample time. Shi \emph{ et al}~\cite{Shiling2011} considered
a system with a single sensor, and studied the problem of whether
or not to send its data to a remote estimator, with the aim of saving
communications. They studied two scheduling schemes, according to
the computational power of the sensor. If the sensor has sufficient
power, and under a given communication constraint, they provide an
optimal scheduling scheme to minimize the mean squared error (MSE).
On the other hand, if the sensor has limited computation power, they
gave a scheduling scheme that guarantees that the MSE remains within
certain prescribed level. Also, an optimal periodic schedule, satisfying
given communication and power constrains, is derived in~\cite{ShilingAC2011}.
Sui \emph{et al}~\cite{sui2015schedule} studied the optimization
of certain sensor scheduling frameworks for the CMSA/CA protocol.
Other relevant works on sensor scheduling include~\cite{hovareshti2007,shiling2007,savage2009,you2013kalman,you2013asym},
to name a few.

When a sensor scheduling scheme is used together with a randomly lossy
data transmission, both scheduling and packet loss directly affect
the stability of estimation process. Our stability results are general
enough so as to be applicable to assess stability in this setup. We
show how this is done using two scheduling schemes, namely, time-based
scheduling and random scheduling.

In order to achieve the above, we generalize the result in~\cite{rohr2014kalman}
in the following senses:
\begin{description}
\item [{Model:}] we drop the diagonalizable assumption on the state transition
matrix $\mathbf{A}$, making the result valid for any arbitrary matrix.
\item [{Measurements:}] we generalized the way in which measurements are
produced in the following three directions:
\begin{enumerate}
\item Instead of considering a random channel model in which a measurement
can be either perfectly transmitted or totally lost, we consider a
far more general random measurement equation, in which, at each sample
time $t$, both the measurement matrix $\mathbf{C}_{t}$ and the measurement
noise covariance $\mathbf{R}_{t}$, are randomly drawn from some known
distribution. 
\item While in current works the most general statistical dependence condition
for the packet loss process is given by the FSMC model, we assume
a more general condition for the pair $\left(\mathbf{C}_{t},\mathbf{R}_{t}\right)$.
This condition is stated in equation~(\ref{eq:independence}).
\item Also, while current works assume that the model for the packet loss
process has stationary statistics, we generalize this assumption to
the case where $\left(\mathbf{C}_{t},\mathbf{R}_{t}\right)$ has cyclo-stationary
statistics. This generalization is essential to the application of
our results to a time-based scheduling setting (Section~\ref{subsec:Time-based-schedule}).
\end{enumerate}
\end{description}
The paper is organized as follows. Section~\ref{sec:Preliminaries}
introduces some mathematical background. Section~\ref{sec:problem}
states the research problem. The main result (Theorem~\ref{thm:main})
is presented in Section~\ref{sec:main}, together with the general
random model of the measurement equation, for which this result is
valid. In Section~\ref{subsec:About-Assumption}, we provide some
insight into this general random model. The stability condition stated
in our main result is expressed in terms of certain quantity, whose
computation is non trivial. In Section~\ref{sec:Phi} we describe
how to compute this quantity. In Section~\ref{sec:schedule}, we
show how to apply our stability results for sensor scheduling. We
draw our conclusions in Section~\ref{sec:conclusion}. For the ease
of reading, the formal proof of our main result is presented in Section~\ref{sec:proof_main}.

\section{Preliminaries\label{sec:Preliminaries}}

Throughout the paper we use the following notation. 
\begin{notation}
\label{nota:common} We use $\mathbb{N}$ to denote the set of natural
numbers, $\mathbb{Z}$ to denote set of integers, $\mathbb{R}$ for
the real numbers and $\mathbb{C}$ for the complex numbers. For a
real or complex scalar, vector or matrix, we use $^{\ast}$ to denote
its transpose conjugate. For an arbitrary set $\mathcal{S}$, we use
$\mathcal{S}^{N}$ to denote the set of $N$-tuples with values in
$\mathcal{S}$, and $\mathcal{S}^{\mathcal{I}}$ for the set of sequences
with the same values indexed by $\mathcal{I}$. For $s\in\mathcal{S}^{N}$
or $s\in\mathcal{S}^{\mathcal{I}}$, we use $s(i)$ to denote the
$i$-th element in $s$.

We use $\mathcal{N}(\mu,\Sigma)$ to denote a normal distribution
with mean $\mu$ and covariance matrix $\Sigma$ and $\mathcal{CN}(\mu,\Sigma)$
to denote a circularly-symmetric complex normal distribution with
the same mean and covariance. For an event $A$, we use $\mathbb{P}\left(A\right)$
to denote its probability. For a random variable $x$, $\mathbb{E}\left(x\right)$
denotes its expectation and $\mathbb{P}\left(x=a\right)$ denotes
the probability of the event $\left\{ x=a\right\} $. Following a
standard convention, in order to simplify the notation, we use $x$
to denote both, the random variable and the value defining the event.
We therefore write $\mathbb{P}\left(x\right)$ as a shorthand notation
for $\mathbb{P}\left(x=x\right)$. 
\end{notation}
We now introduce some required mathematical background on measure
theory.

Let $\left(\mathcal{E},\mathcal{B}\right)$ be a measurable space.
We use $\mathcal{M}(\mathcal{E})$ to denote the Banach space of signed
measures on $\mathcal{E}$, which are bounded in the total variation
norm (which we denote by $\left\Vert \cdot\right\Vert $). We also
use $\mathcal{L}\left(\mathcal{E}\right)$ to denote the set of bounded
linear operators $\kappa:\mathcal{M}\left(\mathcal{E}\right)\rightarrow\mathcal{M}\left(\mathcal{E}\right)$.
We use $\sigma\left(\kappa\right)$ to denote the spectrum of $\kappa\in\mathcal{L}\left(\mathcal{E}\right)$
and $\rho\left(\kappa\right)$ to denote its spectral radius. We also
define $\prod_{n=1}^{N}\kappa_{n}=\kappa_{N}\cdots\kappa_{1}$ and
$\kappa^{N}=\prod_{n=1}^{N}\kappa$. 

An important subset of $\mathcal{M}\left(\mathcal{E}\right)$ is that
of probability measures, which we denote by $\mathcal{P}\left(\mathcal{E}\right)$.
We use $\mathcal{K}\left(\mathcal{E}\right)\subset\mathcal{L}\left(\mathcal{E}\right)$
to denote the set of stochastic transition maps $\kappa:\mathcal{P}(\mathcal{E})\rightarrow\mathcal{P}(\mathcal{E})$.
A stochastic kernel is a map $\boldsymbol{\kappa}:\mathcal{E}\rightarrow\mathcal{P}(\mathcal{E})$
such that, for every $B\in\mathcal{B}$, the map $e\mapsto\boldsymbol{\kappa}(e)(B)$
is measurable. We use $\kappa\in\mathcal{K}\left(\mathcal{E}\right)$
to denote the stochastic transition map induced by $\boldsymbol{\kappa}$
as follows
\[
\kappa\mu\left(B\right)=\int_{\mathcal{E}}\kappa(\varepsilon)(B)\mu\left(d\varepsilon\right).
\]

We finish this section by defining certain elements from the above
spaces, which we will use in the rest of the paper. We define $\delta_{e}\in\mathcal{P}(\mathcal{E})$
by $\delta_{e}(A)=1$ if $e\in A$ and $0$ otherwise. For $D\in\mathcal{B}$
we use $\chi_{D}\in\mathcal{L}\left(\mathcal{E}\right)$ to denote
the map assigning each measure to its restriction to $D$, i.e.,
\[
\left(\chi_{D}\mu\right)(B)=\mu\left(D\cap B\right),\text{ for all }B\in\mathcal{B},\mu\in\mathcal{M}\left(\mathcal{E}\right).
\]
For $\mu\in\mathcal{M}\left(\mathcal{E}\right)$ and $\kappa\in\mathcal{L}\left(\mathcal{E}\right)$,
we define $\mathcal{U}\left(\kappa,\mu\right)\subset\mathcal{M}\left(\mathcal{E}\right)$
to be the set of accumulation points of the sequence $\kappa^{n}\mu$,
i.e., the set of all $\nu\in\mathcal{M}\left(\mathcal{E}\right)$
such that, for every $\epsilon>0$, there exist infinitely many $n\in\mathbb{N}$
such that $\left\Vert \kappa^{n}\mu-\nu\right\Vert $$<\epsilon$.
We define
\begin{equation}
\mathcal{U}\left(\kappa\right)=\overline{\mathrm{span}}(\bigcup_{\mu\in\mathcal{M}\left(\mathcal{E}\right)}\mathcal{U}(\kappa,\mu)),\label{eq:def-U}
\end{equation}
where $\overline{\mathrm{span}}\left(\mathcal{U}\right)$ denotes
the closed linear span of the set $\mathcal{U}$. We also use 
\begin{equation}
\breve{\kappa}=\left.\kappa\right|_{\mathcal{U}\left(\kappa\right)}:\mathcal{U}\left(\kappa\right)\rightarrow\mathcal{U}\left(\kappa\right)\label{eq:breve-opp}
\end{equation}
to denote the restriction of $\kappa$ to $\mathcal{U}\left(\kappa\right)$.
Finally, for a collection $\mathcal{U}\subset\mathcal{M}\left(\mathcal{E}\right)$
of measures, we define the collection of sets $\mathcal{F}\left(\mathcal{U}\right)\subset\mathcal{B}$,
as those which are not null with respect to some measure in $\mathcal{U},$i.e.,
\begin{equation}
\mathcal{F}\left(\mathcal{U}\right)=\left\{ A\in\mathcal{B}:\mu(A)>0\text{ for some }\mu\in\mathcal{U}\right\} .\label{eq:def-F}
\end{equation}

\section{Problem formulation\label{sec:problem}}

Consider the discrete-time linear system 
\begin{align}
\mathbf{x}_{t+1} & =\mathbf{A}\mathbf{x}_{t}+\mathbf{w}_{t},\label{eq:system1}\\
\mathbf{y}_{t} & =\mathbf{C}_{t}\mathbf{x}_{t}+\mathbf{v}_{t},\label{eq:system2}
\end{align}
where $\mathbf{x}_{t}\in\mathbb{C}^{n}$ is the vector of states,
$\mathbf{y}_{t}\in\mathbb{C}^{p}$ is the vector of measurements,
$\mathbf{w}_{t}\sim\mathcal{CN}(\mathbf{0},\mathbf{Q})$, with $\mathbf{Q}\geq0$,
is the process noise, $\mathbf{v}_{t}\sim\mathcal{CN}(\mathbf{0},\mathbf{R}_{t})$
with $\mathbf{R}_{t}\ge0$, is the measurement noise, $\mathbf{A}\in\mathbb{C}^{n\times n}$
is the state matrix and $\mathbf{C}_{t}\in\mathbb{C}^{p\times n}$
is the measurement matrix at time $t$. It is assumed, without loss
of generality, that $\mathbf{A}$ is in Jordan normal form. The initial
state is $\mathbf{x}_{0}\sim\mathcal{CN}(\mathbf{0},\mathbf{P}_{0})$,
with $\mathbf{P}_{0}\geq0$. Also, the set of random vectors $\left\{ \mathbf{x}_{0},\mathbf{w}_{t},\mathbf{v}_{t}:t\geq0\right\} $
is jointly statistically independent. At time $t$, the pair $\gamma_{t}=\left(\mathbf{C}_{t},\mathbf{R}_{t}\right)$
is randomly drawn from the finite set $\mathcal{A}=\mathcal{C}\times\mathcal{R}$,
where $\mathcal{C}=\left\{ \mathbf{C}^{(1)},\cdots,\mathbf{C}^{(D)}\right\} $
and $\mathcal{R}=\left\{ \mathbf{R}^{(1)},\cdots,\mathbf{R}^{(E)}\right\} $.
For $T\in\mathbb{N}$, let $\Gamma_{t,T}=\left(\gamma_{t},\cdots,\gamma_{t+T-1}\right)\in\mathcal{A}^{T}$
denote the random sequence of measurement matrices and noise covariances
from time $t$ up to time $t+T-1$.

Our next step is to introduce the model describing the statistics
of $\gamma_{t}$. We assume that $\gamma_{t}$ is generated by a hidden
Markov model whose state is an element of $\mathcal{E}$. More specifically,
let $h:\mathcal{E}\rightarrow\mathcal{A}$ be a measurable function,
$\mu_{0}\in\mathcal{P}\left(\mathcal{E}\right)$ and $\boldsymbol{\kappa}_{t}:\mathcal{E}\rightarrow\mathcal{P}\left(\mathcal{E}\right)$,
$t\in\mathbb{Z}$, be a sequence of stochastic kernels. The sequence
$\gamma_{t}$ is generated as follows: $\varrho_{0}\sim\mu_{0}$,
and, for each $t>0$,
\begin{align}
\varrho_{t} & \sim\boldsymbol{\kappa}_{t}\left(\varrho_{t-1}\right),\label{eq:netmod1}\\
\gamma_{t} & =h\left(\varrho_{t}\right),\label{eq:netmod2}
\end{align}
where $\varrho\sim\mu$ denotes that $\varrho$ is independently drawn
from the probability distribution $\mu$. We assume that, for each
$s\geq0$, the $\left\{ \mathbf{x}_{0},\mathbf{w}_{t},\mathbf{v}_{t},\varrho_{s}:t\geq0\right\} $
is jointly statistically independent.
\begin{rem}
\label{rem:HMM} We assume the above model for $\gamma_{t}$ without
loss of generality, as it is equivalent to the general model characterized
by specifying $\mathbb{P}\left(\gamma_{t}|\gamma_{s},s<t\right)$,
for all $t\in\mathbb{Z}$ and all possible values of $\gamma_{s}$,
$s\leq t$. To see this, notice that the latter can be written in
the form~(\ref{eq:netmod1})-(\ref{eq:netmod2}) by taking $\mathcal{E}=\mathcal{A}^{\mathbb{N}}$,
$\varrho_{t}=\left(\gamma_{s}:s\leq t\right)$, and defining the $\sigma$-algebra
$\mathcal{B}$ to be the one generated by the sets $C_{t,A}$, $t\in\mathbb{N}$,
$A\subset\mathcal{A}$, where
\[
C_{t,A}=\left\{ \Gamma\in\mathcal{A}^{\mathbb{N}}:\Gamma(t)\in A\right\} .
\]
\end{rem}
A Kalman filter is used to obtain an estimate $\hat{\mathbf{x}}_{t|t-1}$
of the state $\mathbf{x}_{t}$ given the knowledge of $\mathbf{y}_{0},\cdots,\mathbf{y}_{t-1}$
and $\Gamma_{0,t}$. The update equation of the expected covariance
(EC) $\mathbf{P}_{t}=\mathbb{E}\left(\tilde{\mathbf{x}}_{t}\tilde{\mathbf{x}}_{t}^{\ast}\right)$,
with $\tilde{\mathbf{x}}_{t}=\mathbf{x}_{t}-\hat{\mathbf{x}}_{t|t-1}$,
is 
\begin{equation}
\mathbf{P}_{t+1}=\psi_{\gamma_{t}}\left(\mathbf{P}_{t}\right),\label{eq:update_3}
\end{equation}
with 
\begin{multline*}
\psi_{\gamma_{t}}\left(\mathbf{P}_{t}\right)=\\
\mathbf{A}\mathbf{P}_{t}\mathbf{A}^{*}+\mathbf{Q}-\mathbf{A}\mathbf{P}_{t}\mathbf{C}_{t}^{*}\left(\mathbf{C}_{t}\mathbf{P}_{t}\mathbf{C}_{t}^{*}+\mathbf{R}_{t}\right)^{-1}\mathbf{C}_{t}\mathbf{P}_{t}\mathbf{A}^{*}.
\end{multline*}

In this work we derive a necessary condition and a sufficient condition,
with a trivial gap between them (Remark~\ref{rem:trivial-gap} explains
what this means), for the stability of the Kalman filter with a random
measurement equation. This is done by studying the asymptotic norm
of the expected error covariance (ANEEC). In order to define the ANEEC,
we introduce the following notation 
\[
\mathbf{\Psi}\left(\mathbf{P}_{t},\Gamma_{t,T}\right)=\psi_{\gamma_{t+T-1}}\cdots\psi_{\gamma_{t+1}}\psi_{\gamma_{t}}\left(\mathbf{P}_{t}\right),
\]
i.e., $\mathbf{\Psi}\left(\mathbf{P}_{t},\Gamma_{t,T}\right)$ denotes
the covariance matrix resulting at time $t+T$, after starting with
covariance $\mathbf{P}_{t}$ at time $t$, and then applying the sequence
of random measurement equations defined by $\Gamma_{t,T}$. This matrix
depends on the random sequence $\Gamma_{t,T}$ and the initial covariance
$\mathbf{P}_{t}$. In order to work with a quantity independent of
these values, in defining the ANEEC, we take expectation with respect
to $\Gamma_{t,T}$ and the supremum with respect to $\mathbf{P}_{t}$.
This leads to the following definition.
\begin{defn}
\label{def:G} The ANEEC is defined as 
\[
G=\sup_{\begin{subarray}{c}
t\in\mathbb{Z}\\
\mathbf{P}_{t}\geq0
\end{subarray}}\limsup_{T\rightarrow\infty}\left\Vert \mathbb{E}\left(\mathbf{\Psi}\left(\mathbf{P}_{t},\Gamma_{t,T}\right)\right)\right\Vert .
\]
\end{defn}
\begin{rem}
In~(\ref{eq:system2}) we assume that the measurements $\mathbf{y}_{t}\in\mathbb{R}^{p}$
have time-invariant dimension $p$. This assumption is done to simplify
the presentation, and without loss of generality. This is because
the case $\mathbf{y}_{t}\in\mathbb{R}^{p_{t}}$ with time-varying
dimension $p_{t}$ can be handled by defining $p$ as the maximum
number of rows among the matrices $\mathbf{C}^{(d)}$, $d=1,\cdots,D$,
and zero padding the matrices $\mathbf{C}^{(d)}$ and $\mathbf{R}^{(d)}$
so that all of them have $p$ rows.
\end{rem}

\section{Main result\label{sec:main}}

Our main result is stated in terms of certain partition of the system~\eqref{eq:system1}-\eqref{eq:system2}
into subsystems which we call finite multiplicative order (FMO) blocks.
This partition is introduced next. 
\begin{defn}
\label{def:CFMO}A set of complex numbers $x_{i}\in\mathbb{C}$, $i=1,\cdots,I$,
is said to have a common finite multiplicative order $N\in\mathbb{N}$
up to a constant $\alpha\in\mathbb{C}$, if $x_{i}^{N}=\alpha^{N}$,
for all $i=1,\cdots,I$. If there do not exist $N$ and $\alpha$
satisfying the above, the set is said not to have common finite multiplicative
order.
\end{defn}
\begin{exmp}
\label{exa:CFMO}The set of numbers $\left\{ 2,2i,-2,-2i\right\} $
have common finite multiplicative order $4$ up to $2$. The set $\left\{ 1,e^{\sqrt{2}i}\right\} $
does not have common finite multiplicative order.
\end{exmp}
\begin{notation}
\label{nota:1}Consider the following partition of $\mathbf{A}$,
\begin{equation}
\mathbf{A}=\mathrm{diag}(\mathbf{A}_{1},\cdots,\mathbf{A}_{K}),\label{eq:diag_A}
\end{equation}
where the sub-matrices $\mathbf{A}_{k}$ are chosen such that, for
any $k$, the diagonal entries of $\mathbf{A}_{k}$ have a common
finite multiplicative order $N_{k}$ up to $\alpha_{k}$, and for
any $k$ and $l$ with $k\neq l$, the diagonal entries of the matrix
$\mathrm{diag}(\mathbf{A}_{k},\mathbf{A}_{l})$ do not have common
finite multiplicative order.

Let $\bar{J}\in\mathbb{N}$ be the largest size among the Jordan blocks
of $\mathbf{A}$, and $\bar{J}_{k}\in\mathbb{N}$ be the largest size
among the Jordan blocks of $\mathbf{A}_{k}$. For convenience, we
assume that the sub-matrices $\mathbf{A}_{k}$ are ordered such that
$|\alpha_{1}|\geq|\alpha_{2}|\geq\cdots\geq|\alpha_{K}|$. Also, when
$|\alpha_{k}|=|\alpha_{k+1}|$, then $\bar{J}_{k}\geq\bar{J}_{k+1}$.

For each $d=1,\cdots,D$, consider the partition 
\[
\mathbf{C}^{(d)}=\left[\mathbf{C}_{1}^{(d)},\cdots,\mathbf{C}_{K}^{(d)}\right],
\]
such that, for each $k$, $\mathbf{C}_{k}^{(d)}$ has the same number
of columns as $\mathbf{A}_{k}$. Let $\mathcal{C}_{k}=\left\{ \mathbf{C}_{k}^{(d)}:d=1,\cdots,D\right\} $.
\end{notation}
\begin{defn}
\label{def:FMO} In the above partition, each pair $\left(\mathbf{A}_{k},\mathcal{C}_{k}\right)$
is called an FMO block of the system~\eqref{eq:system1}-\eqref{eq:system2}.
\end{defn}
\begin{rem}
Notice that if $\left(\mathbf{A}_{k},\mathcal{C}_{k}\right)$ is an
FMO block, then each sub-matrix $\mathbf{A}_{k}$ can be written as
\begin{eqnarray}
\mathbf{A}_{k} & = & \alpha_{k}\tilde{\mathbf{A}}_{k},\\
\tilde{\mathbf{A}}_{k} & = & \mathrm{diag}\left\{ \exp(i2\pi\theta_{k,1}),\cdots,\exp(i2\pi\theta_{k,K_{k}})\right\} +\mathbf{U}_{k},\label{eq:def_tildeA}
\end{eqnarray}
where $\mathbf{U}_{k}$ is strictly upper triangular, i.e., its non-zero
entries lie above its main diagonal. Also, $\alpha_{k}\in\mathbb{C}$
and $\theta_{k,j}\in\mathbb{Q}$, for $j=1,\cdots,K_{k}$. Notice
that for any $k$ and $l$ with $k\neq l$, $\alpha_{k}/\alpha_{l}$
is not a root of unity, i.e., $(\alpha_{k}/\alpha_{l})^{m}\neq1$
for all $m\in\mathbb{N}$. 
\end{rem}
In stating our main result, we use certain observability matrix ${\bf O}_{k}$
associated to each FMO block $k$ of the system. Our next step is
to introduce this matrix. The measurements $\mathbf{z}_{t,T}^{\ast}=\left[\mathbf{y}_{t}^{\ast},\cdots,\mathbf{y}_{t+T-1}^{\ast}\right]$
available from time $t$ up to $T-1$ can be written as
\begin{equation}
\mathbf{z}_{t,T}=\mathbf{O}\left(\Gamma_{t,T}\right)\mathbf{x}_{t}+\mathbf{f}_{t,T}\left(\Gamma_{t,T}\right),\label{avail_meas}
\end{equation}
where the observability matrix $\mathbf{O}\left(\Gamma_{t,T}\right)$
is given by 
\begin{align}
{\bf O}\left(\Gamma_{t,T}\right) & =\left[\begin{matrix}{\bf O}_{1}\left(\Gamma_{t,T}\right) & {\bf O}_{2}\left(\Gamma_{t,T}\right) & \ldots & {\bf O}_{K}\left(\Gamma_{t,T}\right)\end{matrix}\right],\label{eq:def_O}\\
\mathbf{f}_{t}\left(\Gamma_{t,T}\right) & =\left[\begin{matrix}\mathbf{v}_{t}\\
\mathbf{C}_{t+1}\mathbf{w}_{t}+\mathbf{v}_{t+1}\\
\vdots\\
\mathbf{C}_{T+t-1}\sum_{j=t}^{T+t-2}\mathbf{A}^{T+t-2-j}\mathbf{w}_{j}+\mathbf{v}_{t+T-1}
\end{matrix}\right],\nonumber 
\end{align}
with 
\begin{align*}
{\bf O}_{k}\left(\Gamma_{t,T}\right) & =\left[\begin{matrix}{\bf C}_{t,k}\\
{\bf C}_{t+1,k}{\bf A}_{k}\\
\vdots\\
{\bf C}_{T+t-1,k}{\bf A}_{k}^{T-1}
\end{matrix}\right],\;\mathbf{C}_{t}=\left[\mathbf{C}_{t,1},\cdots,\mathbf{C}_{t,K}\right],
\end{align*}
such that, for each $k$, $\mathbf{C}_{t,k}$ have the same number
of columns as $\mathbf{A}_{k}$.

Our main result is stated in terms of the probability that each matrix
${\bf O}_{k}\left(\Gamma_{t,T}\right)$ does not have full-column
rank (FCR). The following definition identifies the event associated
to sequences leading to this property.
\begin{defn}
For $k=1,\cdots,K$, let 
\begin{equation}
{\cal N}_{k}^{t,T}\triangleq\{\Gamma_{t,T}:\mathbf{O}_{k}\left(\Gamma_{t,T}\right)\text{ does not have FCR}\}.\label{eq:N}
\end{equation}
\end{defn}
We now state our main result. This requires Assumptions~\ref{assumption:1}
and~\ref{assumption:2}. These conditions are rather general. For
this reason, their statement is somewhat technical. In Section~\ref{subsec:About-Assumption}
we give interpretations of these assumptions, as well as more practical
conditions guaranteeing them. However, these assumptions also hold
under conditions more general than those given in Section~\ref{subsec:About-Assumption}.
An example of this appears in the proof of Corollary~\ref{cor:non-deg}.
This shows the value of the generality of Assumptions~\ref{assumption:1}
and~\ref{assumption:2}.
\begin{defn}
\label{def:cyclo} We say that the sequence $\gamma_{t}$, $t\in\mathbb{Z}$,
is cyclostationary with period $\tau\in\mathbb{N}$, if, in~(\ref{eq:netmod1})-(\ref{eq:netmod2}),
we have
\begin{align*}
\kappa_{t} & =\kappa_{t+\tau},\text{ for all }t\in\mathbb{N},\\
\mu_{0} & =\prod_{t=1}^{\tau}\kappa_{t},\mu_{0}.
\end{align*}
We say that it is stationary if it is cyclostationary with $\tau=1$.
\end{defn}
\begin{assumption}
\label{assumption:1} The sequence $\gamma_{t}$, $t\in\mathbb{Z}$,
in~(\ref{eq:netmod1})-(\ref{eq:netmod2}), is cyclostationary with
period $\tau\in\mathbb{N}$ and 
\begin{equation}
\zeta\triangleq\sup_{\begin{subarray}{c}
T\in\mathbb{N}\\
0\leq t<\tau
\end{subarray}}\sup_{\begin{subarray}{c}
\mathbb{P}\left(\varrho_{t-1}\right)\neq0\\
\mathbb{P}\left(\Gamma_{t,T}\right)\neq0
\end{subarray}}\frac{\mathbb{P}\left(\Gamma_{t,T}|\varrho_{t-1}\right)}{\mathbb{P}\left(\Gamma_{t,T}\right)}<\infty.\label{eq:independence}
\end{equation}
\end{assumption}
\begin{assumption}
\label{assumption:2} For any $0\leq t<\tau$, any multiple $M$ of
$\tau$ and any finite collection $\mathcal{D}=\left(D_{m}\in\mathcal{B}:m=1,\cdots,M\right)$,
let
\[
\eta_{t}=\prod_{m=1}^{M}\chi_{D_{m}}\kappa_{t+m}.
\]
Then, 
\begin{equation}
\rho\left(\eta_{t}\right)=\rho\left(\breve{\eta}_{t}\right).\label{eq:restriction}
\end{equation}
Also, for any non-zero non-negative $\mu\in\mathcal{U}\left(\eta_{t}\right)$
and $A\in\mathcal{F}\left(\mathcal{U}\left(\eta_{t}\right)\right)$,
there exists $N$ such that
\begin{equation}
\breve{\eta}_{t}^{n}\mu\left(A\right)>0,\text{ for all }n\geq N.\label{eq:non-support}
\end{equation}
\end{assumption}
We now state our main result. Its proof is deferred to Section~\ref{sec:proof_main}.
\begin{thm}
\label{thm:main} Consider the system~\eqref{eq:system1}-\eqref{eq:system2}
satisfying Assumptions~\ref{assumption:1} and~\ref{assumption:2}.
For $k\in\{1,\cdots,K\}$, let 
\[
\Phi_{k}=\max_{0\leq t<\tau}\limsup_{T\rightarrow\infty}\mathbb{P}\left({\cal N}_{k}^{t,T}\right)^{1/T},
\]
with ${\cal N}_{k}^{t,T}$ defined by~(\ref{eq:N}). If 
\begin{equation}
|\alpha_{k}|^{2}\Phi_{k}<1,\textrm{ for all }k\in\{1,\cdots,K\},\label{eq:if_part}
\end{equation}
then $G<\infty$, and if 
\begin{equation}
|\alpha_{k}|^{2}\Phi_{k}>1,\textrm{ for some }k\in\{1,\cdots,K\},\label{eq:only_if_part}
\end{equation}
then $G=\infty$. 
\end{thm}
\begin{rem}
\label{rem:trivial-gap}Notice that Theorem~\ref{thm:main} is inconclusive
in the case when $|\alpha_{k}|^{2}\Phi_{k}=1.$ Trivial gaps of this
kind are common in the literature~\cite{sinopoli2004kfi,mo2012kfi}. 
\end{rem}

\section{About Assumptions~\ref{assumption:1} and~\ref{assumption:2}\label{subsec:About-Assumption}}

In this section we give an interpretation of the technical condition
stated in Assumptions~\ref{assumption:1} and~\ref{assumption:2}.
We also show that these assumptions hold under certain conditions
which are easier to interpret. This result is given in Proposition~\ref{prop:assumptions}
stated below.
\begin{defn}
A random process $\gamma_{t}$ is Markov of order $L\in\mathbb{N}$
if, for all $m\geq1$, 
\[
\mathbb{P}\left(\gamma_{t}|\gamma_{t-L-m},\cdots,\gamma_{t-1}\right)=\mathbb{P}\left(\gamma_{t}|\gamma_{t-L},\cdots,\gamma_{t-1}\right).
\]
Furthermore, it is called proper if all the above probabilities are
strictly bigger than zero. Finally, the process is independent if
$L=0$.
\end{defn}
\begin{defn}
A random process $\gamma_{t}$ is called Gaussian hidden Markov if
it is generated by a hidden Markov model like~(\ref{eq:netmod1})-(\ref{eq:netmod2}),
but with~(\ref{eq:netmod1}) replaced by 
\[
\varrho_{t}=\mathbf{K}\varrho_{t-1}+\varepsilon_{t},
\]
where $\mathbf{K}$ is a stable matrix (i.e., $\rho\left(\mathbf{K}\right)<1$)
and $\varepsilon_{t}\sim\mathcal{N}\left(0,\Sigma\right)$, with $\left\{ \mathbf{x}_{0},\mathbf{w}_{t},\mathbf{v}_{t},\varepsilon_{t}:t\geq0\right\} $
being a jointly independent set of random vectors.
\end{defn}
\begin{prop}
\label{prop:assumptions} Suppose that $\gamma_{t}$ is cyclostationary
with period $\tau$ and is either finite-order proper Markov, or Gaussian
hidden Markov. Then Assumptions~\ref{assumption:1} and~\ref{assumption:2}
hold.
\end{prop}
The reminder of this section is devoted to show Proposition~\ref{prop:assumptions}.

If $\gamma_{t}$ is an independent sequence, we could simply take
$\mathcal{E}=\mathcal{A}$, $\varrho_{t}=\gamma_{t}$ and $h$ to
be the identity map. Then,~(\ref{eq:independence}) would hold trivially.
Hence,~(\ref{eq:independence}) can be interpreted as a generalization
of the independence property. More generally, the following two lemmas
provide conditions under which~(\ref{eq:independence}) holds without
the independence property.
\begin{lem}
\label{lem:zeta1} If $\gamma_{t}$ is cyclostationary with period
$\tau\in\mathbb{N}$ and finite-order Markov, then~(\ref{eq:independence})
holds. 
\end{lem}
\begin{pf}
Since $\gamma_{t}$ is finite order Markov, we can take $\varrho_{t}=\Gamma_{t-L+1,L}$.
We then have 
\begin{align*}
\zeta & =\sup_{\begin{subarray}{c}
T\in\mathbb{N}\\
0\leq t<\tau
\end{subarray}}\sup_{\begin{subarray}{c}
\mathbb{P}\left(\varrho_{t-1}\right)\neq0\\
\mathbb{P}\left(\Gamma_{t,T}\right)\neq0
\end{subarray}}\frac{\mathbb{P}\left(\Gamma_{t,T}|\varrho_{t-1}\right)}{\mathbb{P}\left(\Gamma_{t,T}\right)}\\
 & =\sup_{\begin{subarray}{c}
T\in\mathbb{N}\\
0\leq t<\tau
\end{subarray}}\sup_{\begin{subarray}{c}
\mathbb{P}\left(\varrho_{t-1}\right)\neq0\\
\mathbb{P}\left(\Gamma_{t,T}\right)\neq0
\end{subarray}}\frac{\mathbb{P}\left(\varrho_{t-1}|\Gamma_{t,T}\right)}{\mathbb{P}\left(\varrho_{t-1}\right)}\\
 & \leq\sup_{0\leq t<\tau}\sup_{\mathbb{P}\left(\varrho_{t-1}\right)\neq0}\frac{1}{\mathbb{P}\left(\varrho_{t-1}\right)}\\
 & <\infty,
\end{align*}
where the last inequality follows since the supremum operations are
taken over finite sets. 
\end{pf}
\begin{lem}
\label{lem:zeta2} If $\gamma_{t}$ is cyclostationary with period
$\tau$ and Gaussian hidden Markov, then~(\ref{eq:independence})
holds.
\end{lem}
\begin{pf}
Let $\mathcal{E}=\mathbb{R}^{d}$ and $\varrho_{t}\in\mathcal{E}$.
Then there exist a partition $\mathcal{S}=\left\{ \mathcal{S}^{(d,e)}:d=1,\cdots,D,e=1,\cdots,E\right\} $
of $\mathbb{R}^{d}$ (i.e., $\mathcal{S}^{(d,e)}\cap\mathcal{S}^{(d^{\prime},e^{\prime})}=\emptyset$,
whenever $(d,e)\neq(d^{\prime},e^{\prime})$, and $\bigcup_{(d,e)=(1,1)}^{(D,E)}\mathcal{S}^{(d,e)}=\mathbb{R}^{d}$),
such that 
\[
h(\varrho)=\left(\mathbf{C}^{(d)},\mathbf{E}^{(e)}\right),\text{ for all }\varrho\in\mathcal{S}^{(d,e)}.
\]
Let $\mathcal{S}_{t}=h^{-1}\left(\gamma_{t}\right)$, i.e., the unique
element from the partition $\mathcal{S}$ such that $h\left(\varrho\right)=\gamma_{t}$,
for all $\varrho\in\mathcal{S}_{t}$, and let $\mathcal{S}_{t,T}=\mathcal{S}_{t}\times\cdots\times\mathcal{S}_{t+T-1}$.
We have that
\begin{equation}
\mathbb{P}\left(\varrho_{t-1}|\Gamma_{t,T}\right)=\mathbb{P}\left(\varrho_{t-1}|\varrho_{t,T}\in\mathcal{S}_{t,T}\right).\label{eq:puapua}
\end{equation}
The right-hand side of~(\ref{eq:puapua}) can be considered as the
probability of $\varrho_{t-1}$ conditioned on the future output of
a stationary quantizer. Since the process $\varrho_{t}$ is Gaussian,
it is easy but tedious to show that there exists a constant $c>0$
such that 
\[
\frac{\mathbb{P}\left(\varrho_{t-1}|\varrho_{t,T}\in\mathcal{S}_{t,T}\right)}{\mathbb{P}\left(\varrho_{t-1}\right)}<c,\text{ for all }t,T,\varrho_{t-1},\text{ and }\mathcal{S}_{t,T}.
\]
We then have
\begin{align*}
\zeta & =\sup_{\begin{subarray}{c}
T\in\mathbb{N}\\
0\leq t<\tau
\end{subarray}}\sup_{\begin{subarray}{c}
\mathbb{P}\left(\varrho_{t-1}\right)\neq0\\
\mathbb{P}\left(\Gamma_{t,T}\right)\neq0
\end{subarray}}\frac{\mathbb{P}\left(\Gamma_{t,T}|\varrho_{t-1}\right)}{\mathbb{P}\left(\Gamma_{t,T}\right)}\\
 & =\sup_{\begin{subarray}{c}
T\in\mathbb{N}\\
0\leq t<\tau
\end{subarray}}\sup_{\begin{subarray}{c}
\mathbb{P}\left(\varrho_{t-1}\right)\neq0\\
\mathbb{P}\left(\Gamma_{t,T}\right)\neq0
\end{subarray}}\frac{\mathbb{P}\left(\varrho_{t-1}|\Gamma_{t,T}\right)}{\mathbb{P}\left(\varrho_{t-1}\right)}\\
 & =\sup_{\begin{subarray}{c}
T\in\mathbb{N}\\
0\leq t<\tau
\end{subarray}}\sup_{\mathcal{S}_{t,T},\varrho_{t-1}}\frac{\mathbb{P}\left(\varrho_{t-1}|\varrho_{t,T}\in\mathcal{S}_{t,T}\right)}{\mathbb{P}\left(\varrho_{t-1}\right)}\\
 & \leq c<\infty.
\end{align*}
\end{pf}
We now turn our attention to condition~(\ref{eq:non-support}). In
this condition, the output of the probability transition mapping $\kappa_{t}$
is restricted to a set $D_{m}$ (via the restriction operator $\chi_{D_{m}}$),
which is taken form a finite family $\mathcal{D}$ of sets. The resulting
operators are then composed by sequentially taking all sets within
$\mathcal{D}$. This yields the map $\eta_{t}$. The set $\mathcal{\mathcal{U}}\left(\eta_{t}\right)$
of measures is invariant under $\eta_{t}$. Condition~(\ref{eq:restriction})
requires that the spectrum of $\eta_{t}$ equals that of its restriction
to its invariant subspace $\mathcal{U}\left(\eta_{t}\right)$. Also,
$\mathcal{F}\left(\mathcal{U}\left(\eta_{t}\right)\right)$ contains
all sets which are non-null for some measure in $\mathcal{U}\left(\eta_{t}\right)$.
Condition~(\ref{eq:non-support}) requires that the measure of all
sets in $\mathcal{F}\left(\mathcal{U}\left(\eta_{t}\right)\right)$
become eventually and persistently strictly positive, when, starting
from any measure in $\mathcal{U}\left(\eta_{t}\right)$, we sequentially
apply the map $\eta_{t}$.

The above condition seems in principle difficult to verify. However,
the two lemmas below show that it holds under conditions similar to
those in Lemmas~\ref{lem:zeta1} and~\ref{lem:zeta2}.
\begin{lem}
\label{lem:nonsupport1} If $\gamma_{t}$ is finite-order proper Markov,
then Assumption~\ref{assumption:2} holds.
\end{lem}
\begin{pf}
Recall the definition of $\eta_{t}$ given in Assumption~\ref{assumption:2}.
Let L be the Markov order of $\gamma_{t}$. Let $\mathcal{E}=\mathcal{A}^{L}$
and $\varrho_{t}=\Gamma_{t-L+1,L}$. It is easy to see that $\mathcal{F}\left(\mathcal{U}\left(\eta_{t}\right)\right)\subseteq\mathcal{G}$
where
\[
\mathcal{G}\triangleq\left\{ \varrho\in\mathcal{E}:\varrho(l)\subset D_{M-\mathrm{mod}\left(L-l,M\right)},\lambda(A)>0\right\} .
\]
Now, for any $\mu$ and $n>L/M$, the measure $\eta_{t}^{n}\mu$ is
strictly positive on any $A\in\mathcal{G}$. Hence,~(\ref{eq:non-support})
holds. Also, $\eta_{t}^{n}\mu(\varrho)=0$, for all $\varrho\notin\mathcal{G}$.
Hence,~(\ref{eq:restriction}) also holds and the result follows.
\end{pf}
\begin{lem}
\label{lem:nonsupport2} If $\gamma_{t}$ is Gaussian hidden Markov,
then Assumption~\ref{assumption:2} holds.
\end{lem}
\begin{pf}
Recall the definition of $\eta_{t}$ given in Assumption~\ref{assumption:2}.
It is easy to see that 
\[
\mathcal{F}\left(\mathcal{U}\left(\eta_{t}\right)\right)=\left\{ A\in\mathcal{B}:A\subset D_{M},\lambda(A)>0\right\} .
\]
Also, for any $\mu$, $\eta_{t}^{n}\mu$ has density $g_{n}$ with
respect to the Lebesgue's measure $\lambda$, and this density is
$\lambda$-almost everywhere strictly positive on $D_{M}$. Hence,
for all $A\in\mathcal{F}\left(\mathcal{U}\left(\eta_{t}\right)\right)$,
\[
\eta_{t}^{n}\mu(A)=\int_{A}g_{n}d\lambda>0,\text{ for all }n\geq1,
\]
and~(\ref{eq:non-support}) holds. Also, $\eta_{t}\mu(A)=0$ for
all $A\in\mathcal{B}$ satisfying $\lambda\left(B\cap D_{M}\right)=0$.
So~(\ref{eq:restriction}) holds and the result follows.
\end{pf}
\begin{pf}
{[}of Proposition~\ref{prop:assumptions}{]} It follows by combining
Lemmas~\ref{lem:zeta1},~\ref{lem:zeta2},~\ref{lem:nonsupport1}
and~\ref{lem:nonsupport2}.
\end{pf}

\section{Computing $\Phi_{k}$\label{sec:Phi}}

Our result in Theorem~\ref{thm:main} is stated in terms of the quantities
$\Phi_{k}$, $k=1,\cdots,K$. We introduce below results on how to
compute this quantity. For the easy of readability, their proofs are
deferred to the appendix. Since in our study the choice of $k=1,\cdots,K$
is fixed, to remove $k$ from the notation, we consider a generic
FMO block $\left(\mathbf{A},\mathcal{C}\right)$. We start by introducing
some necessary notation.
\begin{notation}
\label{nota:2} Let $N\in\mathbb{N}$ be the smallest positive integer
such that $\mathbf{A}^{N}=\alpha^{N}\mathbf{I}$. Let $\mathbb{K}=\left\{ \mathrm{ker}\left(\mathbf{O}\left(\Gamma\right)\right):\Gamma\in\mathcal{A}^{N}\right\} \cup\left\{ \mathbb{C}^{n},\emptyset\right\} $
be the set of all possible kernels of $\mathbf{O}\left(\Gamma\right)$,
for sequences $\Gamma$ of length $N$, including, the whole space
$\mathbb{C}^{n}$ and the empty set $\emptyset$. Notice that $\mathbb{K}$
includes all possible kernels of $\mathbf{O}\left(\Gamma\right)$
for sequences $\Gamma$ of length $nN$, for any $n\in\mathbb{N}$.
For any $n\in\mathbb{N}$, define the map $\psi:\mathcal{A}^{nN}\rightarrow\mathbb{K}$
by 
\[
\psi\left(\Gamma\right)=\mathrm{ker}\left(\mathbf{O}(\Gamma)\right).
\]
Let $\mathcal{I}=\{0,\cdots,I\}$ and $\mathcal{K}_{i}$, $i\in\mathcal{I}$
denote all the elements in $\mathbb{K}$. The elements $\mathcal{K}_{i}$
are numerated such that, if $\mathcal{K}_{i}\subset\mathcal{K}_{j}$,
then $i>j$ (notice that, in particular, $\mathcal{K}_{0}=\mathbb{C}^{n}$
and $\mathcal{K}_{I}=\emptyset$). Let $M\in\mathbb{N}$ be any common
multiple of $N$ and $\tau$. For each $t\in\mathbb{N}_{0}$, $i,j\in\mathcal{I}$,
$e\in\mathcal{E}$, $A\in\mathcal{B}$, and $\pi\in\mathcal{L}\left(\mathcal{E}\right)$,
let $T_{t}:\mathcal{I}\times\mathcal{I}\times\mathcal{E}\times\mathcal{B}\rightarrow[0,1]$
be defined by 
\[
T_{t}\left(i,j,e,A\right)=\mathbb{P}\left(\varrho_{t+M}\in A,\psi\left(\Gamma_{t,M}\right)\cap\mathcal{K}_{j}=\mathcal{K}_{i}|\varrho_{t}=e\right),
\]
and $\varsigma_{t}:\mathcal{I}\times\mathcal{I}\rightarrow\mathcal{L}\left(\mathcal{E}\right)$
by
\[
\varsigma_{t}(i,j)\pi(A)=\int T_{t}\left(i,j,e,A\right)\pi(de).
\]
\end{notation}
The next result provides a method for evaluating $\Phi$.
\begin{prop}
\label{prop:deg} Let $\left(\mathbf{A},\mathcal{C}\right)$ be an
FMO block. If Assumptions~\ref{assumption:1} and~\ref{assumption:2}
hold, then 
\begin{equation}
\Phi=\max_{0\leq t<\tau}\max_{0\leq i<I}\rho\left(\breve{\varsigma}_{t}(i,i)\right)^{1/M}.\label{eq:alg_claim}
\end{equation}
\end{prop}
The above result requires computing the spectral radius of the map
$\breve{\varsigma}_{t}(i,i)$. If $\mathcal{E}$ is a discrete finite
space, then $\breve{\varsigma}_{t}(i,i)$ becomes a matrix and $\rho\left(\breve{\varsigma}_{t}(i,i)\right)$
can be easily computed. Otherwise, the following result can be used.
\begin{prop}
\label{prop:spec-rad}For every non-zero non-negative $\mu\in\mathcal{U}\left(\varsigma_{t}(i,i)\right)$,
\[
\rho\left(\breve{\varsigma}_{t}(i,i)\right)=\lim_{n\rightarrow\infty}\left\Vert \breve{\varsigma}_{t}^{n}(i,i)\mu\right\Vert ^{1/n}.
\]
\end{prop}
The following corollary states how the expression~(\ref{eq:alg_claim})
simplifies in the particular case when there exists a single measurement
matrix $\mathbf{C}^{(\alpha)}$ producing measurements which would
never make the observability matrix have FCR (e.g., when measurements
are lost), but any single measurement produced by any other matrix
$\mathbf{C}^{(d)},$ $d\neq\alpha$, would.
\begin{cor}
\label{cor:non-deg} If Assumption~\ref{assumption:1} holds, there
exists $1\leq\alpha\leq D$ such that pair $\left(\mathbf{A},\mathbf{C}^{(\alpha)}\right)$
is non-observable, for each $d\neq\alpha$, $\mathbf{C}^{(d)}$ has
FCR, and there exists $1\leq\beta\leq E$ such that $\mathbb{P}\left(\mathbf{R}_{t}|\mathbf{C}_{t}=\mathbf{C}^{(e)}\right)=\delta_{\mathbf{R}^{(\beta)}}$
(i.e., there is only one possible value of $\mathbf{R}_{t}$ compatible
with $\mathbf{C}_{t}=\mathbf{C}^{(\alpha)}$), then 
\begin{equation}
\Phi=\prod_{t=0}^{\tau-1}\mathbb{P}\left(\mathbf{C}_{t}=\mathbf{C}^{(\alpha)}|\mathbf{C}_{s}=\mathbf{C}^{(\alpha)},s<t\right)^{1/\tau}.\label{eq:claim33}
\end{equation}
\end{cor}

\section{Application: Sensor scheduling with packet loss\label{sec:schedule}}

We have a linear system whose dynamics is given by
\begin{equation}
\mathbf{p}_{t+1}=\mathbf{F}\mathbf{p}_{t}+\mathbf{n}_{t},\label{eq:app0}
\end{equation}
with $\mathbf{n}_{t}^{(s)}\sim\mathcal{N}\left(0,\mathbf{N}\right)$.
There are $S$ sensors. For each $s=1,\cdots,S$, sensor $s$ measures
\begin{equation}
\mathbf{u}_{t}^{(s)}=\mathbf{H}^{(s)}\mathbf{p}_{t}+\mathbf{e}_{t}^{(s)},\label{eq:app1}
\end{equation}
with $\mathbf{e}_{t}^{(s)}\sim\mathcal{N}\left(0,\mathbf{E}^{(s)}\right)$.
We assume that measurements from only $R<S$ sensors can be transmitted
at each time instant $t$. Then, at each time we transmit 
\[
\mathbf{r}_{t}=\left(\mathbf{M}_{t}\otimes\mathbf{I}\right)\mathbf{u}_{t},
\]
where $\mathbf{u}_{t}^{T}=\left[\left(\mathbf{u}_{t}^{(1)}\right)^{T},\cdots,\left(\mathbf{u}_{t}^{(S)}\right)^{T}\right]$
and $\mathbf{M}_{t}$ is the row-selection matrix determining the
schedule at time $t$. Since there are packet losses, the actual transmission
is given by 
\begin{equation}
\mathbf{y}_{t}=\left(\mathbf{L}_{t}\otimes\mathbf{I}\right)\mathbf{r}_{t},\label{eq:app2}
\end{equation}
where $\mathbf{L}_{t}=\mathrm{diag}\left\{ l_{t}^{(1)},\cdots,l_{t}^{(R)}\right\} $
and $l_{t}^{(r)}$ is a binary random variable determining whether
the packet associated with the $r$-th scheduled measurement was lost
($l_{t}^{(r)}=0$) or not ($l_{t}^{(r)}=1$). Let $\mathbf{A}=\mathbf{V}\mathbf{F}\mathbf{V}^{-1}$
be the Jordan normal form of $\mathbf{F}$. Then, the system equations
are given by~\eqref{eq:system1}-\eqref{eq:system2}, with 
\[
\begin{array}{ccc}
\mathbf{C}_{t}=\mathbf{B}_{t}\mathbf{H}\mathbf{V^{-1}}, & \mathbf{Q}_{t}=\mathbf{V}\mathbf{N\mathbf{V}}^{T}, & \mathbf{R}_{t}=\mathbf{B}_{t}\mathbf{E}\mathbf{B}_{t}^{T},\\
\mathbf{x}_{t}=\mathbf{V}\mathbf{p}_{t}, & \mathbf{w}_{t}=\mathbf{V}\mathbf{n}_{t}, & \mathbf{v}_{t}=\mathbf{B}_{t}\mathbf{e}_{t},
\end{array}
\]
and 
\begin{align*}
\mathbf{e}_{t}^{T} & =\left[\left(\mathbf{e}_{t}^{(1)}\right)^{T},\cdots,\left(\mathbf{e}_{t}^{(S)}\right)^{T}\right],\\
\mathbf{H}^{T} & =\left[\left(\mathbf{H}^{(1)}\right)^{T},\cdots,\left(\mathbf{H}^{(S)}\right)^{T}\right],\\
\mathbf{E} & =\mathrm{diag}\left\{ \mathbf{E}^{(1)},\cdots,\mathbf{E}^{(S)}\right\} ,\\
\mathbf{B}_{t} & =\left(\mathbf{L}_{t}\otimes\mathbf{I}\right)\left(\mathbf{M}_{t}\otimes\mathbf{I}\right).
\end{align*}

We consider below two scheduling strategy, namely, time-based schedule
and random schedule.

\subsection{Time-based schedule\label{subsec:Time-based-schedule}}

In this case, the packet loss model $\mathbf{L}_{t}$ is stationary,
independent of $\mathbf{w}_{t}$, $\mathbf{v}_{t}$ and $\mathbf{x}_{0}$,
and either, finite-order proper Markov or Gaussian hidden Markov.
The sequence of matrices $\mathbf{M}_{t}$, $t\in\mathbb{N}$, follows
a periodic deterministic pattern, i.e., 
\[
\mathbf{M}_{t}=\mathbf{M}_{t+\tau},
\]
for all $t\in\mathbb{N}$ and some period $\tau\in\mathbb{N}$. Clearly,
this leads to $\gamma_{t}$ being cyclostationary with period $\tau$,
and satisfying the conditions of Proposition~\ref{prop:assumptions}.
Theorem~\ref{thm:main} then holds.

\subsection{Random schedule}

In this case, both sequence of matrices $\mathbf{L}_{t}$ and $\mathbf{M}_{t}$
are randomly drawn at each $t\in\mathbb{N}$. This is done such that
the sequences $\left(\mathbf{M}_{t}\right)$, $\left(\mathbf{L}_{t}\right)$,
$\left(\mathbf{w}_{t}\right)$, $\left(\mathbf{v}_{t}\right)$ and
$\mathbf{x}_{0}$ are mutually independent. The models describing
the statistics of $\mathbf{L}_{t}$ and $\mathbf{M}_{t}$ are stationary
and either finite-order proper Markov or Gaussian hidden Markov. This
clearly leads to $\gamma_{t}=\left(\mathbf{C}_{t},\mathbf{E}_{t}\right)$
satisfying the conditions of Proposition~\ref{prop:assumptions}.
Theorem~\ref{thm:main} thus holds.

\subsection{Example\label{subsec:Example}}

In this section we use Theorem~\ref{thm:main} and Corollary~\ref{cor:non-deg}
to assess the stability of an example system. We consider a system
whose state-transition matrix $\mathbf{A}$ is non-diagonalizable
and whose measurement equation have statistics which are not finite-order
Markov. Notice that, as mentioned in points~1 and~3 in the introduction,
none of the results available in the literature could be used to assess
the stability of a system with any of these two properties.

Consider a system whose dynamics is given by~\eqref{eq:system1},
with 
\[
\mathbf{A}=\mathrm{diag}\left\{ \mathbf{A}_{1},\mathbf{A}_{2}\right\} ,\;\mathbf{A}_{1}=\left[\begin{array}{cc}
\alpha_{1} & 1\\
0 & \alpha_{1}
\end{array}\right],\:\mathbf{A}_{2}=\left[\alpha_{2}\right],
\]
for some $\alpha_{1}>\alpha_{2}>0$. There are two sensors. For $i\in\{1,2\}$,
the measurement equation of sensor $i$ is given by~(\ref{eq:app1}),
with 
\[
\mathbf{H}^{(1)}=\left[\begin{array}{ccc}
2 & 1 & 0\\
0 & 1 & 0
\end{array}\right]\text{ and }\mathbf{H}^{(2)}=\left[\begin{array}{ccc}
0 & 0 & 1\\
0 & 0 & 2
\end{array}\right],
\]
Due to communication constraints, the measurements from both sensors
are alternatively transmitted, i.e., 
\[
\mathbf{M}_{t}=\begin{cases}
\left[\begin{array}{cc}
\mathbf{I} & 0\end{array}\right] & t\text{ even},\\
\left[\begin{array}{cc}
0 & \mathbf{I}\end{array}\right] & t\text{ odd}.
\end{cases}
\]
We assume that the communication channel has a packet loss model given
by $\mathbf{L}_{t}=l_{t}\in\left\{ 0,1\right\} $, where $l_{t}$
is Gaussian hidden Markov. Hence, we have a time-based schedule, as
described in Section~\ref{subsec:Time-based-schedule}. Thus, we
can use the result of Theorem~\ref{thm:main} to determine the stability
of the Kalman filter.

We have 
\[
\boldsymbol{\mathbf{B}}_{t}=\begin{cases}
\left[\begin{array}{cc}
l_{t}\mathbf{I} & 0\end{array}\right] & t\text{ even},\\
\left[\begin{array}{cc}
0 & l_{t}\mathbf{I}\end{array}\right] & t\text{ odd}.
\end{cases}
\]
Hence, the measurement equation of the aggregated system is given
by~(\ref{eq:system2}), with $\mathbf{C}_{t}$ given by
\begin{equation}
\mathbf{\boldsymbol{C}}_{t}=\begin{cases}
l_{t}\left[\begin{array}{ccc}
2 & 1 & 0\\
0 & 1 & 0
\end{array}\right] & t\text{ even},\\
l_{t}\left[\begin{array}{ccc}
0 & 0 & 1\\
0 & 0 & 2
\end{array}\right] & t\text{ odd}.
\end{cases}\label{eq:ex_C}
\end{equation}

From Definition~\ref{def:FMO} the FMO blocks of the above system
are $\left(\mathbf{A}_{1},\mathcal{C}_{1}\right)$ and $\left(\mathbf{A}_{2},\mathcal{C}_{2}\right)$
where
\[
\mathcal{C}_{k}=\left\{ \mathbf{C}_{k}^{(1)},\mathbf{C}_{k}^{(2)}\right\} ,\,k=1,2,
\]
with
\[
\begin{array}{cc}
\mathbf{C}_{1}^{(1)}=\left[\begin{array}{cc}
0 & 0\\
0 & 0
\end{array}\right], & \mathbf{C}_{1}^{(2)}=\left[\begin{array}{cc}
2 & 1\\
0 & 1
\end{array}\right],\\
\mathbf{C}_{2}^{(1)}=\left[\begin{array}{c}
0\\
0
\end{array}\right], & \mathbf{C}_{2}^{(2)}=\left[\begin{array}{c}
1\\
2
\end{array}\right].
\end{array}
\]
Clearly, both FMO blocks satisfy the conditions of Corollary~\ref{cor:non-deg}.
Hence, we can apply this result to each block. Let 
\[
\lambda=\mathbb{P}\left(l_{t}=0|l_{2s}=0,s<t\right).
\]
We have
\begin{align*}
\mathbb{P}\left(\mathbf{C}_{t,1}=\mathbf{C}_{1}^{(0)}|\mathbf{C}_{s,1}=\mathbf{C}_{1}^{(0)},s<t\right) & =\begin{cases}
\lambda & t\text{ even},\\
1 & t\text{ odd},
\end{cases}\\
\mathbb{P}\left(\mathbf{C}_{t,2}=\mathbf{C}_{2}^{(0)}|\mathbf{C}_{s,2}=\mathbf{C}_{2}^{(0)},s<t\right) & =\begin{cases}
1 & t\text{ even},\\
\lambda & t\text{ odd},
\end{cases}
\end{align*}
Then, from~(\ref{eq:claim33}), since the cyclostationary period
is $\tau=2$, we obtain
\[
\Phi_{1}=\Phi_{2}=\lambda^{1/2}.
\]
It then follows from Theorem~\ref{thm:main} that 
\begin{align*}
\alpha_{1}^{4}\lambda<1 & \Rightarrow G<\infty,\\
\alpha_{1}^{4}\lambda>1 & \Rightarrow G=\infty.
\end{align*}

\section{Proof of the main result\label{sec:proof_main}}

This section presents a formal proof of the necessary and the sufficient
conditions stated in Theorem~\ref{thm:main}. In Section~\ref{subsec:Preliminary-results}
we derive certain preliminary results. More precisely, in Section~\ref{subsec:Bounds}
we provide lower and upper bounds on the growth rate of $\|\mathbf{\Psi}\left(\mathbf{P}_{t},\Gamma_{t,T}\right)\|$,
and in Section~\ref{subsec:kernel-orth} we state a technical condition
to guarantee that the kernel of $\mathbf{O}\left(\Gamma_{t,T}\right)$
has certain desired orientation. In Section~\ref{sec:nec_cond} we
show the necessary condition. In Section~\ref{subsec:A-first-sufficient}
we derive a first sufficient condition, which differs from the desired
one. This result is used in Section~\ref{sec:second_suff_cond} to
provide a second sufficient condition, seemingly stronger than the
one in Theorem~\ref{thm:main}. We then show in Section~\ref{subsec:equivalence}
that the latter condition is indeed equivalent to the desired one.

\subsection{Preliminary results\label{subsec:Preliminary-results}}

\subsubsection{Bounds on the growth rate of $\|\mathbf{\Psi}\left(\mathbf{P}_{t},\Gamma_{t,T}\right)\|$\label{subsec:Bounds}}

It turns out that the growth rate of $\|\mathbf{\Psi}\left(\mathbf{P}_{t},\Gamma_{t,T}\right)\|$
is determined by the location of the kernel of $\mathbf{O}\left(\Gamma_{t,T}\right)$.
Recall from~\eqref{avail_meas} that 
\begin{align}
\mathbf{z}_{T} & =\mathbf{O}\left(\Gamma_{t,T}\right)\mathbf{x}_{t}+\mathbf{f}_{t}\left(\Gamma_{t,T}\right)\label{eq:Y}\\
\mathbf{x}_{T} & =\mathbf{A}^{T-t}\mathbf{x}_{t}+\mathbf{q}_{t,T}\label{eq:xT}\\
\mathbf{q}_{t,T} & =\sum_{j=t}^{t+T-1}\mathbf{A}^{t+T-1-j}\mathbf{w}_{j}.\label{eq:q_t}
\end{align}
From~\cite[Ch. 5, Theorem 2.1]{anderson1979of}, we have 
\begin{equation}
\mathbf{\Psi}\left(\mathbf{P}_{t},\Gamma_{t,T}\right)=\mathbf{\Sigma}_{\mathbf{x}}-\mathbf{\Sigma}_{\mathbf{x},\mathbf{z}}\mathbf{\Sigma}_{\mathbf{z}}^{\dagger}\mathbf{\Sigma}_{\mathbf{x},\mathbf{z}}^{*},\label{eq:PT}
\end{equation}
where $^{\dagger}$ is the Moore-Penrose pseudo-inverse~\cite{ben2003git}
and 
\begin{align*}
\mathbf{\Sigma}_{\mathbf{x}} & =\mathbf{A}^{T}\mathbf{P}_{t}\mathbf{A}^{*T}+\mathbb{E}\left(\mathbf{q}_{t,T}\mathbf{q}_{t,T}^{*}\right),\\
\mathbf{\Sigma}_{\mathbf{z}} & =\mathbf{O}\left(\Gamma_{t,T}\right)\mathbf{P}_{t}\mathbf{O}\left(\Gamma_{t,T}\right)^{\ast}+\mathbb{E}\left(\mathbf{f}_{t}\left(\Gamma_{t,T}\right)\mathbf{f}\left(\Gamma_{t,T}\right)^{\ast}\right),\\
\mathbf{\Sigma}_{\mathbf{x},\mathbf{z}} & =\mathbf{A}^{T}\mathbf{P}_{t}\mathbf{O}\left(\Gamma_{t,T}\right)^{\ast}+\mathbb{E}\left(\mathbf{q}_{t,T}\mathbf{f}_{t}\left(\Gamma_{t,T}\right)^{\ast}\right).
\end{align*}

\begin{lem}
\label{lem:norm_bnd} Let $\mathbf{A}$ be a Jordan block of order
$J$ with eigenvalue $\alpha$. Then, there exist $c_{1},c_{2}\in\mathbb{R}$
such that 
\begin{align}
\|\mathbf{A}^{t}\| & \leq|\alpha|^{t}c_{1}t^{J-1}\label{eq:bound1}\\
\left\Vert \mathbf{A}^{-t}\right\Vert ^{-1} & \geq|\alpha|^{t}c_{2}t^{1-J}\label{eq:bound2}
\end{align}
for all $t\in\mathbb{N}$. 
\end{lem}
\begin{pf}
The proof is divided in two steps.

Proof of~(\ref{eq:bound1}): Notice that 
\[
\mathbf{A}^{t}=\alpha^{t}\mathbf{M}(t)
\]
where 
\[
\mathbf{M}(t)=\left[\begin{matrix}1 & p_{1}(t) & \cdots & p_{J-1}(t)\\
0 & 1 & \ddots & p_{J-2}(t)\\
\vdots & \ddots & \ddots & \vdots\\
0 & 0 & \cdots & 1
\end{matrix}\right],
\]
and $p_{j}(t)$ is a polynomial in $t$ of order $j$ given by 
\[
p_{j}(t)={t \choose j}\alpha^{-j}.
\]
We then have 
\begin{align*}
\left\Vert \mathbf{A}^{t}\right\Vert  & =|\alpha|^{t}\|\mathbf{M}(t)\|\\
 & \overset{\text{(a)}}{\leq}|\alpha|^{t}\left(1+\sum_{j=1}^{J-1}|p_{j}(t)|\right)\\
 & \leq|\alpha|^{t}c_{1}t^{J-1},
\end{align*}
for some $c_{1}\in\mathbb{R}$, where~(a) follows from Young's inequality~\cite[p. 115]{Katznelson200401}.

Proof of~(\ref{eq:bound2}): Consider the matrix 
\[
\tilde{\mathbf{M}}(t)=\left[\begin{matrix}1 & \tilde{p}_{1}(t) & \cdots & \tilde{p}_{J-1}(t)\\
0 & 1 & \ddots & \tilde{p}_{J-2}(t)\\
\vdots & \ddots & \ddots & \vdots\\
0 & 0 & \cdots & 1
\end{matrix}\right],
\]
where $\tilde{p}_{j}(t)$ are polynomials in $t$ of order $j$ such
that $\tilde{\mathbf{M}}(t)=\mathbf{M}^{-1}(t)$. This is always possible,
since 
\[
\mathbf{M}(t)\tilde{\mathbf{M}}(t)=\left[\begin{matrix}1 & \hat{p}_{1}(t) & \cdots & \hat{p}_{J-1}(t)\\
0 & 1 & \ddots & \hat{p}_{J-2}(t)\\
\vdots & \ddots & \ddots & \vdots\\
0 & 0 & \cdots & 1
\end{matrix}\right],
\]
with 
\[
\hat{p}_{j}(t)=p_{j}(t)+\tilde{p}_{j}(t)+\sum_{i=1}^{j-1}p_{j-i}(t)\tilde{p}_{i}(t).
\]
From the above, by making 
\[
\tilde{p}_{j}(t)=-p_{j}(t)-\sum_{i=1}^{j-1}p_{j-i}(t)\tilde{p}_{i}(t),
\]
we have $\mathbf{M}(t)\tilde{\mathbf{M}}(t)=\mathbf{I}$. Then, 
\begin{align}
\left\Vert \mathbf{A}^{-t}\right\Vert  & =\|\alpha^{-t}\mathbf{M}^{-1}(t)\|\\
 & =|\alpha|^{-t}\|\tilde{\mathbf{M}}(t)\|\\
 & \leq|\alpha|^{-t}\left(1+\sum_{j=1}^{J-1}|\tilde{p}_{j}(t)|\right)\\
 & \leq|\alpha|^{-t}c_{3}t^{J-1},
\end{align}
for some $c_{3}\in\mathbb{R}$. Hence, 
\[
\left\Vert \mathbf{A}^{-t}\right\Vert ^{-1}\geq|\alpha|^{t}c_{3}^{-1}t^{1-J}.
\]
The result follows by making $c_{2}=c_{3}^{-1}$. 
\end{pf}
The following two lemmas state bounds on the growth rate of $\|\mathbf{\Psi}\left(\mathbf{P}_{t},\Gamma_{t,T}\right)\|$.
Firstly, some notation is introduced. Let $\mathbf{e}_{k,j}$ be the
column vector with a $1$ in one entry and zeros otherwise, such that
$\mathbf{e}_{k,j}^{\top}\mathbf{A}\mathbf{e}_{k,j}$ equals the $j$-th
diagonal entry of the $k$-th block $\mathbf{A}_{k}$ of $\mathbf{A}$.
Let also $\mathcal{E}_{k}=\{\mathbf{e}_{k,1},\cdots,\mathbf{e}_{k,K_{k}}\}$.
The following lemma states an upper bound on the growth rate of $\|\mathbf{\Psi}\left(\mathbf{P}_{t},\Gamma_{t,T}\right)\|$. 
\begin{lem}
\label{lem:upper-bound} Consider the system~\eqref{eq:system1}-\eqref{eq:system2}.
If $\ker\left\{ \mathbf{O}\left(\Gamma_{t,T}\right)\right\} \subseteq\mathrm{span}\{\mathcal{E}_{k},\cdots,\mathcal{E}_{K}\}$
for some $1\leq k\leq K$, then, there exist $l_{T}>0$ and $c_{1}>0$,
such that, for any $\mathbf{P}_{t}$, 
\begin{equation}
\|\mathbf{\Psi}\left(\mathbf{P}_{t},\Gamma_{t,T}\right)\|\leq|\alpha_{k}|^{2t}c_{1}T^{2(\bar{J}-1)}\|\mathbf{P}_{t}\|+l_{T}.\label{eq:upper-bound1}
\end{equation}
\end{lem}
Also, if $\ker\left\{ \mathbf{O}\left(\Gamma_{t,T}\right)\right\} =\{\mathbf{0}\}$,
then 
\begin{equation}
\|\mathbf{\Psi}\left(\mathbf{P}_{t},\Gamma_{t,T}\right)\|\leq l_{T}.\label{eq:upper-bound2}
\end{equation}

\begin{pf}
Following the steps of the proof of~\cite[Lemma 20]{rohr2014kalman},
we obtain (\ref{eq:upper-bound2}) and 
\begin{align}
\mathbf{\Psi}\left(\mathbf{P}_{t},\Gamma_{t,T}\right) & \leq\mathbf{M}_{t,T}+l_{T}\mathbf{I},\label{eq:aux1}
\end{align}
where $\mathbf{M}_{t,T}=\mathbf{\Pi}\mathbf{A}^{T}\mathbf{P}_{t}\mathbf{A}^{*T}\mathbf{\Pi}$
and 
\[
\mathbf{\Pi}=\mathrm{diag}(\mathbf{0}_{1},\cdots,\mathbf{0}_{k-1},\mathbf{X}),
\]
with $\mathbf{0}_{j}$ being a square matrix of zeros with the same
dimension of $\mathbf{A}_{j}$ and $\mathbf{X}$ is a non-zero matrix
with appropriate dimensions. Also, 
\[
l_{T}=\sup_{t\in\mathbb{Z}}\max_{\Gamma_{t,T}\in\mathcal{A}^{T}}\vartheta_{t}\left(\Gamma_{t,T}\right),
\]
with
\begin{align*}
\vartheta_{t}\left(\Gamma_{t,T}\right) & =\left\Vert \mathbb{E}\left(\mathbf{U}_{t}\left(\Gamma_{t,T}\right)\mathbf{U}_{t}^{\ast}\left(\Gamma_{t,T}\right)\left|\Gamma_{t,T}\right.\right)\right\Vert ,\\
\mathbf{U}_{t}\left(\Gamma_{t,T}\right) & =\mathbf{q}_{t,T}-\left(\mathbf{O}\left(\Gamma_{t,T}\right)\mathbf{A}^{-T}\right)^{\dagger}\mathbf{f}_{t}\left(\Gamma_{t,T}\right).
\end{align*}
Notice that the map $\vartheta_{t}:\mathcal{A}^{T}\rightarrow\mathbb{R}$
is independent of $t\in\mathbb{Z}$. Hence, $l_{T}$ is the result
of a maximization over the finite set $\mathcal{A}^{T}$. This guarantees
that $l_{T}$ is finite. Hence,~(\ref{eq:upper-bound2}) clearly
holds.

Let $\mathbf{M}_{t,T}=\mathbf{N}_{t,T}^{*}\mathbf{N}_{t,T}$, with
$\mathbf{N}_{t,T}=\mathbf{P}_{t}^{1/2}\mathbf{A}^{*T}\mathbf{\Pi}$
and $\tilde{\mathbf{A}}=\mathrm{diag}\{\mathbf{0},\cdots,\mathbf{0},\mathbf{A}_{k},\cdots,\mathbf{A}_{K}\}$.
Then, 
\begin{align*}
\|\mathbf{M}_{t,T}\| & =\|\mathbf{N}_{t,T}\|^{2}\\
 & \leq\|\mathbf{P}_{t}\|\|\mathbf{A}^{*T}\mathbf{\Pi}\|^{2}\\
 & =\|\mathbf{P}_{t}\|\|\tilde{\mathbf{A}}^{*T}\mathbf{\Pi}\|^{2}\\
 & \leq\|\mathbf{P}_{t}\|\|\tilde{\mathbf{A}}^{*T}\|^{2}\|\mathbf{\Pi}\|^{2}.
\end{align*}
Notice that $\|\tilde{\mathbf{A}}^{*T}\|=\max_{j\geq k}\|\mathbf{A}_{j}^{T}\|$.
Using Lemma~\ref{lem:norm_bnd}, we obtain 
\[
\|\mathbf{M}_{t,T}\|\leq\|\mathbf{P}_{t}\||\alpha_{k}|^{2T}c_{1}T^{2(\bar{J}-1)},
\]
for some $c_{1}\in\mathbb{R}$ and the result follows from~\eqref{eq:aux1}. 
\end{pf}
The next lemma states a lower bound on the growth rate of $\|\mathbf{\Psi}\left(\mathbf{P}_{t},\Gamma_{t,T}\right)\|$. 
\begin{lem}
\label{lem:lower-bound} Consider the system~\eqref{eq:system1}-\eqref{eq:system2}.
If $\ker\left\{ \mathbf{O}\left(\Gamma_{t,T}\right)\right\} \cap\mathrm{span}\{\mathcal{E}_{k}\}\neq\{0\}$,
then, there exists $c_{2}>0$ such that for all $\mathbf{P}_{t}$,
\[
\left\Vert \mathbf{\Psi}\left(\mathbf{P}_{t},\Gamma_{t,T}\right)\right\Vert \geq|\alpha_{k}|^{2T}c_{2}T^{2(1-\bar{J})}\left\Vert \mathbf{P}_{t}^{-1}\right\Vert ^{-1}.
\]
\end{lem}
\begin{pf}
Following the steps of the proof of~\cite[Lemma 21]{rohr2014kalman},
we obtain 
\begin{equation}
\left\Vert \mathbf{\Psi}\left(\mathbf{P}_{t},\Gamma_{t,T}\right)\right\Vert \geq\left\Vert \mathbf{P}_{t}^{-1}\right\Vert ^{-1}\left\Vert \mathbf{A}^{T}\mathbf{\Pi}\right\Vert ^{2},\label{eq:subs_normP}
\end{equation}
where $\mathbf{\Pi}=\left[\mathbf{I}-\mathbf{O}\left(\Gamma_{t,T}\right)^{\dagger}\mathbf{O}\left(\Gamma_{t,T}\right)\right]$.
Now, let $\mathbf{x}\in\ker\left\{ \mathbf{O}\left(\Gamma_{t,T}\right)\right\} \cap\mathrm{span}\{\mathcal{E}_{k}\}$.
Since $\mathbf{\Pi}$ is the projection onto the kernel of $\mathbf{O}\left(\Gamma_{t,T}\right)$,
we have 
\[
\left\Vert \mathbf{A}^{T}\mathbf{\Pi}\mathbf{x}\right\Vert =\left\Vert \mathbf{A}^{T}\mathbf{x}\right\Vert .
\]
Since $\mathbf{x}\in\mathrm{span}({\cal E}_{k})$, we have that 
\begin{align}
\|\mathbf{A}^{T}\mathbf{x}\| & =\|\mathbf{A}_{k}^{T}\mathbf{v}\|\\
 & \geq\|\mathbf{A}_{k}^{-T}\|^{-1}\|\mathbf{v}\|\\
 & =\|\mathbf{A}_{k}^{-T}\|^{-1}\|\mathbf{x}\|,
\end{align}
for some $\mathbf{v}\in\mathbb{C}^{N_{k}}$. From Lemma~\ref{lem:norm_bnd},
and the above, it follows that 
\[
\|\mathbf{A}^{T}\mathbf{\Pi}\mathbf{x}\|\geq c_{2}|\alpha_{k}|^{T}T^{1-\bar{J}}\|\mathbf{x}\|,
\]
hence 
\begin{equation}
\|\mathbf{A}^{T}\mathbf{\Pi}\|\geq|\alpha_{k}|^{T}T^{1-\bar{J}}.\label{eq:rep_ch3ah}
\end{equation}
The result then follows by substituting~\eqref{eq:rep_ch3ah} into~\eqref{eq:subs_normP}. 
\end{pf}

\subsubsection{A condition to guarantee that $\ker\{\mathbf{O}(\Gamma_{t,T})\}$
is orthogonal to $\mathrm{span}\{\mathcal{E}_{k}\}$\label{subsec:kernel-orth}}
\begin{defn}
A matrix $\mathbf{M}$ is said to have full column rank with strength
$q\in\mathbb{N}_{0}$ ($\mathbb{N}_{0}=\mathbb{N}\cup\{0\}$), denoted
by $\mathrm{FCR}(q)$, if $\mathbf{M}$ has more than $q$ rows and
the matrix obtained after removing any $q$ rows from $\mathbf{M}$
still has FCR. 
\end{defn}
The main goal of this section is to show the following lemma. 
\begin{lem}
\label{lem:num_th} There exists $Q\in\mathbb{N}_{0}$ such that,
for any $1\leq k\leq K$, if $\mathbf{O}_{k}(\Gamma_{t,T})$ has $\mathrm{FCR}(Q)$,
then $\ker\{\mathbf{O}(\Gamma_{t,T})\}\perp\mathrm{span}\{\mathcal{E}_{k}\}$. 
\end{lem}
The proof of Lemma~\ref{lem:num_th} uses a number of results which
are stated below. 
\begin{defn}
\label{def:apif}{[}Almost periodic function~\cite[p. 45]{corduneanu1989apf}{]}
A function $f:\mathbb{N}\rightarrow\mathbb{C}$ of an integer variable
is called almost periodic, if to any $\epsilon>0$ there corresponds
an integer $N(\epsilon)$, such that among any $N(\epsilon)$ consecutive
integers there exists a $p$ with the property 
\[
|f(n+p)-f(n)|<\epsilon,~n\in\mathbb{N}.
\]
The same definition holds for functions $f:\mathbb{R}\rightarrow\mathbb{C}$
of a real variable, by making $N(\epsilon)$ real, and replacing the
block of $N(\epsilon)$ consecutive integers by an interval of length
$N(\epsilon)$. 
\end{defn}
\begin{lem}
\label{lem:apif}{[}Theorem 1.27 from~\cite{corduneanu1989apf}{]}
A necessary and sufficient condition for $f:\mathbb{N}\rightarrow\mathbb{C}$
to be almost periodic is the existence of an almost periodic function
$g(x)$, $x\in\mathbb{R}$ such that $f(n)=g(n)$, $n\in\mathbb{N}$. 
\end{lem}
\begin{lem}
\label{lem:sml}{[}Skolem Mahler Lech Theorem~\cite{lech1953nrs}{]}
Consider the sequence $s_{n}\in\mathbb{C}$, $n\in\mathbb{N}$, satisfying
the recursion formula 
\begin{equation}
s_{n}=k_{1}s_{n-1}+k_{2}s_{n-2}+\cdots+k_{p}s_{n-p},~n\geq p,\label{eq:recursion2}
\end{equation}
with $k_{i}\in\mathbb{C}$. If $s_{n}=0$ for infinitely many values
of $n$, then those $s_{n}$ that are equal to zero occur periodically
in the sequence from a certain index on. 
\end{lem}
\begin{lem}
\label{lem:evertse} {[}Immediate consequence of~\cite[Theorem 1.2]{evertse2002lev}{]}
Let $0\neq\alpha_{k},b_{k}\in\mathbb{C}$, $k=1,\cdots,K$, with $(\alpha_{k}/\alpha_{j})^{t}\neq1$
for all $k\neq j$ and all $t\in\mathbb{N}$. Then, there exists a
finite number of non-negative integers $t\in\mathbb{N}_{0}$ such
that 
\begin{equation}
\sum_{k=1}^{K}b_{k}\alpha_{k}^{t}=0.\label{eq:main}
\end{equation}
\end{lem}
\begin{lem}
\label{lem:jordan} Let $0\neq\alpha_{k}\in\mathbb{C}$, $k=1,\cdots,K$,
with $(\alpha_{k}/\alpha_{j})^{t}\neq1$ for all $k\neq j$ and all
$t\in\mathbb{N}$. Each $\alpha_{k}$ has an associated $\bar{J}_{k}\in\mathbb{N}_{0}$.
Let the $\alpha$'s be ordered such that $|\alpha_{k}|\geq|\alpha_{k+1}|$
and $\bar{J}_{k}\geq\bar{J}_{k+1}$ whenever $|\alpha_{k}|=|\alpha_{k+1}|$.
Let $c_{k,j}\in\mathbb{C},~k=1,\cdots,K$, $j=1,\cdots,\bar{J}_{k}$,
with at least one $c_{k,j}\neq0$. Then, there exists a finite number
of non-negative integers $t\in\mathbb{N}_{0}$ such that 
\[
f(t)\triangleq\sum_{k=1}^{K}\sum_{j=0}^{\bar{J}_{k}}c_{k,j}\alpha_{k}^{t}t^{j}=0.
\]
\end{lem}
\begin{pf}
The proof is divided in steps.

1) Notice that $f(t)$ can be written as a linear recursion like~\eqref{eq:recursion2}.
Hence, from Lemma~\ref{lem:sml}, it follows that either: 
\begin{itemize}
\item[a)] $f(t)=0$ for a finite number of $t\in\mathbb{N}$, or 
\item[b)] there exist $t_{1},t_{2}$ such that $f(t_{1}+t_{2}k)=0$ for all
$k\in\mathbb{N}$. 
\end{itemize}
Hence, we need to show that b) cannot hold.

2) Let $b$ be the largest $k$ such that $|\alpha_{k}|=|\alpha_{1}|$
and $\bar{J}_{k}=\bar{J}_{1}$. We have that 
\[
\frac{f(t)}{\alpha_{1}^{t}t^{\bar{J}_{1}}}=g(t)+h(t),
\]
with 
\begin{align}
g(t) & =\sum_{k=1}^{b}c_{k,\bar{J}_{1}}\left(\frac{\alpha_{k}}{\alpha_{1}}\right)^{t},\label{eq:def_g}\\
h(t) & =\sum_{k=1}^{b}\sum_{j=0}^{\bar{J}_{k}-1}c_{k,j}\frac{\alpha_{k}^{t}}{\alpha_{1}^{t}}t^{j-\bar{J}_{1}}+\sum_{k=b+1}^{K}\sum_{j=0}^{\bar{J}_{k}}c_{k,j}\frac{\alpha_{k}^{t}}{\alpha_{1}^{t}}t^{j-\bar{J}_{1}}.\nonumber 
\end{align}
Notice that 
\[
\lim_{t\rightarrow\infty}h(t)=0.
\]
Hence, it is enough to show that 
\begin{equation}
\limsup_{t\rightarrow\infty}|g(t)|>0.\label{eq:weneed}
\end{equation}

3) Let $t,T\in\mathbb{N}$ and 
\[
\tilde{g}_{t,T}(n)=g(t+nT),
\]
for all $n\in\mathbb{N}$. From~\eqref{eq:def_g}, we have 
\[
\tilde{g}_{t,T}(n)=\sum_{k=1}^{b}a_{k}\theta_{k}^{n},
\]
with $a_{k}=c_{k,\bar{J}_{1}}\left(\frac{\alpha_{k}}{\alpha_{1}}\right)^{t}$
and $\theta_{k}=\left(\frac{\alpha_{k}}{\alpha_{1}}\right)^{T}$.
Notice that $|\theta_{k}|=1$, for all $k$. Now, for $x\in\mathbb{R}$,
$\tilde{g}_{t,T}(x)$ is a trigonometric polynomial, and therefore
an almost periodic function of a continuous variable~\cite[p. 9]{corduneanu1989apf}.
Then, from Lemma~\ref{lem:apif}, $\tilde{g}_{t,T}(n)$ is an almost
periodic function of an integer variable. It then follows from Definition~\ref{def:apif}
that, for any $n\in\mathbb{N}$ and $\epsilon>0$, there exists an
infinite increasing sequence of integers $p_{l}$, $l\in\mathbb{N}$
such that 
\[
|\tilde{g}_{t,T}(n)-\tilde{g}_{t,T}(n+p_{l})|<\epsilon,\text{ for all }l\in\mathbb{N}.
\]
Hence, 
\begin{equation}
\limsup_{n\rightarrow\infty}|\tilde{g}_{t,T}(n)|=\sup_{n\in\mathbb{Z}}|\tilde{g}_{t,T}(n)|.\label{eq:piopio}
\end{equation}
From Lemma~\ref{lem:evertse}, $\tilde{g}_{t,T}(n)=0$ can only hold
for a finite number of $n$'s. Hence 
\[
\sup_{k\in\mathbb{Z}}|\tilde{g}_{t,T}(k)|>0.
\]
It then follows from~(\ref{eq:piopio}) that, 
\[
\limsup_{k\rightarrow\infty}|g(t+kT)|>\epsilon,
\]
for some $\epsilon>0$. Therefore,~(\ref{eq:weneed}) holds and the
result follows. 
\end{pf}
We can now show Lemma~\ref{lem:num_th}. 
\begin{pf}
{[}of Lemma~\ref{lem:num_th}{]} Recall from~\eqref{eq:def_tildeA}
that $\mathbf{A}_{k}\in\mathbb{C}^{K_{k}\times K_{k}}$ and that for
all $k\neq j$ and all $t\in\mathbb{N}$, $\alpha_{k}^{t}/\alpha_{j}^{t}\neq1$.
Let $\mathbf{v}\in\mathbb{C}^{n}$ be in the kernel of $\mathbf{O}(\Gamma_{t,T})$,
i.e., 
\begin{equation}
\mathbf{O}(\Gamma_{t,T})\mathbf{v}=\mathbf{0},\label{eq:OSvz}
\end{equation}
with $\mathbf{v}=[\mathbf{v}_{1}^{T},\cdots,\mathbf{v}_{K}^{T}]^{T}$
and $\mathbf{v}_{k}\in\mathbb{C}^{K_{k}}$. To show the result it
is enough to show that if $\mathbf{O}_{k}(\Gamma_{t,T})$ has $\mathrm{FCR}(Q)$,
then $\mathbf{v}_{k}=\mathbf{0}$. This is done by contradiction in
four steps.

1) Let $\mathbf{v}_{k}\neq\mathbf{0}$. Let $\Gamma_{t,T}$ be such
that $\mathbf{O}_{k}(\Gamma_{t,T})$ has $\mathrm{FCR}(Q)$. From~\eqref{eq:def_tildeA},
it follows that 
\[
\mathbf{A}_{k}^{s}=\mathrm{diag}\left\{ \mathbf{A}_{k,1}^{s},\cdots,\mathbf{A}_{k,\bar{J}_{k}}^{s}\right\} ,
\]
where $\mathbf{A}_{k,j}$ is a Jordan block. The entry on the $r$-th
row and $c$-th column of $\mathbf{A}_{k,1}^{s}$ is given by 
\begin{multline*}
\left[\mathbf{A}_{k,j}^{s}\right]_{r,c}\\
=\begin{cases}
0, & c<r\\
\alpha_{k}^{s-n}\exp\left(2\pi i\theta_{k,j}(s-n)\right){t \choose n}, & c=r+n,n\geq0.
\end{cases}
\end{multline*}
Let $[\mathbf{C}_{k}^{(d)}]_{r}$ be the $r$-th row of $\mathbf{C}_{k}^{(d)}$
and $[\mathbf{C}^{(d)}]_{r}$ be the $r$-th row of $\mathbf{C}^{(d)}$.
Notice that there exists scalars $c_{k,r,j}^{(d)}(s)$, $1\leq k\leq K$,
$1\leq r\leq p$, $1\leq j\leq\bar{J}_{k}$, $1\leq d\leq D$, such
that 
\[
[\mathbf{C}_{k}^{(d)}]_{r}\mathbf{A}_{k}^{s}\mathbf{v}_{k}=\alpha_{k}^{s}\sum_{j=1}^{\bar{J}_{k}}s^{j-1}c_{k,r,j}^{(d)}(s).
\]
Let $\mathbf{c}_{k,r}^{(d)}(s)=\left[c_{k,r,1}^{(d)}(s),\cdots,c_{k,r,\bar{J}_{k}}^{(d)}(s)\right]$
and $\mathbf{c}_{r}^{(d)}(s)=\left[\mathbf{c}_{1,r}^{(d)}(s),\cdots,\mathbf{c}_{K,r}^{(d)}(s)\right]$.
Recall that $N_{k}$, $k=1,\cdots,K$, are such that all the diagonal
entries of $\mathbf{A}_{k}^{N_{k}}$ are identical. Let $N$ be the
least common multiple of $N_{1},\cdots,N_{K}$. We then have 
\[
\mathbf{c}_{r}^{(d)}(s)=\mathbf{c}_{r}^{(d)}(s+N).
\]
Hence, for given $d$ and $r$, there exists a set ${\cal S}_{r}^{(d)}$,
with $N$ elements, such that for all $s\in\mathbb{N}$ 
\[
\mathbf{c}_{r}^{(d)}(s)\in{\cal S}_{r}^{(d)}.
\]

2) Define 
\[
{\cal U}_{k,r}^{(d)}=\left\{ s\in\mathbb{N}:\mathbf{C}_{s}=\mathbf{C}^{(d)},[\mathbf{C}_{k}^{(d)}]_{r}\mathbf{A}_{k}^{s}\mathbf{v}_{k}\neq0\right\} .
\]
Since $\mathbf{O}_{k}(\Gamma_{t,T})$ has $\mathrm{FCR}(Q)$, there
exist $d$ and $r$ such that the set ${\cal U}_{k,r}^{(d)}$ has
at least $Q/(pD)$ elements.

3) For $n=0,\cdots,N-1$, define the set 
\[
{\cal V}_{k,r,n}^{(d)}=\left\{ s\in{\cal U}_{k,r}^{(d)}:s~\mathrm{mod}~N=n\right\} .
\]
Notice that for all $s,\tilde{s}\in{\cal V}_{k,r,n}^{(d)}$, we have
$c_{k,r,j}^{(d)}(s)=c_{k,r,j}^{(d)}(\tilde{s})$. We then define 
\[
c_{k,r,j,n}^{(d)}\triangleq c_{k,r,j}^{(d)}(s),\,s\in{\cal V}_{k,r,n}^{(d)}.
\]
Also, for at least one $j$, we have $c_{k,r,j,n}^{(d)}\neq0$. For
$s\in{\cal V}_{k,r,n}^{(d)}$ it follows that 
\[
[\mathbf{C}^{(d)}]_{r}\mathbf{A}^{s}\mathbf{v}=\sum_{k=1}^{K}\sum_{j=1}^{\bar{J}_{k}}c_{k,r,j,n}^{(d)}\alpha_{k}^{t}s^{j}.
\]

4) From Lemma~\ref{lem:jordan}, there exist only $E$ values of
$s\in\mathbb{N}$ that satisfy 
\begin{equation}
[\mathbf{C}]_{r}\mathbf{A}^{s}\mathbf{v}=0,\label{eq:CAVZ}
\end{equation}
where $E$ is a finite non-negative integer. Hence, for $Q=(NE+1)pD$,
the set ${\cal U}_{k,r}^{(d)}$ has at least $NE+1$ elements. This
implies that there exists $0\leq n\leq N-1$ such that the set ${\cal V}_{k,r,n}^{(d)}$
has at least $E+1$ elements. Therefore,~\eqref{eq:CAVZ} cannot
hold for all $s\in{\cal V}_{k,r,n}^{(d)}$, implying that $\mathbf{O}(\Gamma_{t,T})\mathbf{v}\neq0$.
This contradiction implies that $\mathbf{v}_{k}$ must be $\mathbf{0}$
in order for~\eqref{eq:OSvz} to hold. 
\end{pf}

\subsection{Proof of the necessary condition in Theorem~\ref{thm:main}\label{sec:nec_cond}}

Following the steps of the proof in~~\cite[Section V-B]{rohr2014kalman}
we obtain 
\begin{multline*}
\left\Vert \mathbb{E}\left(\mathbf{\Psi}\left(\mathbf{P}_{t},\Gamma_{t,T}\right)\right)\right\Vert \geq\\
\max_{1\leq k\leq K}\frac{1}{n}\sum_{\Gamma_{t,T}\in{\cal N}_{k}^{t,T}}\mathbb{P}\left(\Gamma_{t,T}\right)\left\Vert \mathbf{\Psi}\left(\mathbf{P}_{t},\Gamma_{t,T}\right)\right\Vert .
\end{multline*}
From Lemma~\ref{lem:lower-bound}, we have that, for all $k=1,\cdots,K$
and $t\in\mathbb{N}$, 
\begin{align*}
 & \left\Vert \mathbb{E}\left(\mathbf{\Psi}\left(\mathbf{P}_{t},\Gamma_{t,T}\right)\right)\right\Vert \\
\geq & \frac{1}{n}\sum_{\Gamma_{t,T}\in{\cal N}_{k}^{t,T}}\mathbb{P}\left(\Gamma_{t,T}\right)|\alpha_{k}|^{2T}c_{2}T^{2(1-\bar{J})}\left\Vert \mathbf{P}_{t}^{-1}\right\Vert ^{-1}\\
= & \frac{1}{n}|\alpha_{k}|^{2T}c_{2}T^{2(1-\bar{J})}\left\Vert \mathbf{P}_{t}^{-1}\right\Vert ^{-1}\mathbb{P}\left({\cal N}_{k}^{t,T}\right)\\
= & \left(|\alpha_{k}|^{2}\mathbb{P}\left({\cal N}_{k}^{t,T}\right)^{1/T}T^{2(1-\bar{J})/T}\right)^{T}c_{2}\frac{\left\Vert \mathbf{P}_{t}^{-1}\right\Vert ^{-1}}{n}.
\end{align*}
For any $t\in\mathbb{N}$, put $\mathbf{P}_{t}=\mathbf{P}_{0}$. Then,
\begin{align*}
 & \max_{0\leq t<\tau}\limsup_{T\rightarrow\infty}\left\Vert \mathbb{E}\left(\mathbf{\Psi}\left(\mathbf{P}_{t},\Gamma_{t,T}\right)\right)\right\Vert \\
\geq & \max_{0\leq t<\tau}\limsup_{T\rightarrow\infty}\left(|\alpha_{k}|^{2}\mathbb{P}\left({\cal N}_{k}^{t,T}\right)^{1/T}T^{2(1-\bar{J})/T}\right)^{T}\times\\
 & \times c_{2}\frac{\left\Vert \mathbf{P}_{0}^{-1}\right\Vert ^{-1}}{n}\\
= & c_{2}\frac{\left\Vert \mathbf{P}_{0}^{-1}\right\Vert ^{-1}}{n}\max_{0\leq t<\tau}\limsup_{T\rightarrow\infty}a_{t,T}^{T},
\end{align*}
where 
\[
a_{t,T}=|\alpha_{k}|^{2}\mathbb{P}\left({\cal N}_{k}^{t,T}\right)^{1/T}T^{2(1-\bar{J})/T}.
\]

Choose $k=1,\cdots,K$ satisfying~\eqref{eq:only_if_part}. Then
\begin{align*}
 & \max_{0\leq t<\tau}\limsup_{T\rightarrow\infty}a_{t,T}\\
= & \max_{0\leq t<\tau}\limsup_{T\rightarrow\infty}|\alpha_{k}|^{2}\mathbb{P}\left({\cal N}_{k}^{t,T}\right)^{1/T}T^{2(1-\bar{J})/T}\\
= & |\alpha_{k}|^{2}\max_{0\leq t<\tau}\limsup_{T\rightarrow\infty}\mathbb{P}\left({\cal N}_{k}^{t,T}\right)^{1/T}\lim_{T\rightarrow\infty}T^{2(1-\bar{J})/T}\\
= & |\alpha_{k}|^{2}\Phi_{k}\left(\lim_{T\rightarrow\infty}T^{1/T}\right)^{2(\bar{J}-1)}\\
= & |\alpha_{k}|^{2}\Phi_{k}\\
> & 1.
\end{align*}
Hence, if we choose $\left\Vert \mathbf{P}_{0}^{-1}\right\Vert ^{-1}>0$,
then 
\[
G=\max_{0\leq t<\tau}\limsup_{T\rightarrow\infty}\left\Vert \mathbb{E}\left(\mathbf{\Psi}\left(\mathbf{P}_{t},\Gamma_{t,T}\right)\right)\right\Vert =\infty,
\]
and the result follows.

\subsection{Proof of the sufficient condition in Theorem~\ref{thm:main}}

\subsubsection{First step\label{subsec:A-first-sufficient}}

Define the map $\xi_{t,T}:\mathbb{R}_{0}^{+}\to\mathbb{R}_{0}^{+}$
by 
\begin{multline}
\xi_{t,T}(x)=\sup_{\varrho_{t-1}\in\mathcal{E}}\sum_{\Gamma_{t,T}\in\mathcal{A}^{T}}\mathbb{P}\left(\Gamma_{t,T}|\varrho_{t-1}\right)\times\\
\times\mathrm{Tr}\left\{ \mathbf{\Psi}(x\mathbf{I},\Gamma_{t,T})\right\} .\label{eq:def_xi_3}
\end{multline}

\begin{lem}
\label{lem:xi}Let $T,S\in\mathbb{N}$ and $x,y>0$. Then 
\end{lem}
\begin{enumerate}
\item if $x\geq y$, then $\xi_{t,T}(x)\geq\xi_{t,T}(y)$; 
\item $\xi_{t,T+S}(x)\leq\xi_{t+T,S}\circ\xi_{t,T}(x)$.
\end{enumerate}
\begin{pf}
Proof of 1) From~\cite[Lemma 1c]{sinopoli2004kfi}, 
\begin{equation}
x\geq y\Rightarrow\mathbf{\Psi}(x\mathbf{I},\Gamma_{t,T})\geq\mathbf{\Psi}(y\mathbf{I},\Gamma_{t,T}).\label{eq:aux-3}
\end{equation}
Then 
\begin{align*}
\xi_{t,T}(y) & =\sup_{\varrho_{t-1}}\sum_{\Gamma_{t,T}}\mathbb{P}\left(\Gamma_{t,T}|\varrho_{t-1}\right)\mathrm{Tr}\left\{ \mathbf{\Psi}(y\mathbf{I},\Gamma_{t,T})\right\} \\
 & \leq\sup_{\varrho_{t-1}}\sum_{\Gamma_{t,T}}\mathbb{P}\left(\Gamma_{t,T}|\varrho_{t-1}\right)\mathrm{Tr}\left\{ \mathbf{\Psi}(x\mathbf{I},\Gamma_{t,T})\right\} \\
 & =\xi_{t,T}(x).
\end{align*}

Proof of 2) We have 
\begin{align*}
 & \xi_{t,T+S}(x)\\
= & \sup_{\varrho_{t-1}}\sum_{\Gamma_{t,T+S}}\mathbb{P}\left(\Gamma_{t,T+S}|\varrho_{t-1}\right)\times\\
 & \times\mathrm{Tr}\left\{ \mathbf{\Psi}\left(x\mathbf{I},\Gamma_{t,T+S}\right)\right\} \\
= & \sup_{\varrho_{t-1}}\sum_{\Gamma_{t,T}}\sum_{\Gamma_{t+T,S}}\mathbb{P}\left(\Gamma_{t+T,S}|\Gamma_{t,T},\varrho_{t-1}\right)\times\\
 & \mathbb{P}\left(\Gamma_{t,T}|\varrho_{t-1}\right)\mathrm{Tr}\left\{ \mathbf{\Psi}\left(\mathbf{\Psi}\left(x\mathbf{I},\Gamma_{t,T}\right),\Gamma_{t+T,S}\right)\right\} \\
\leq & \sup_{\begin{array}{c}
\varrho_{t+T-1}\\
\varrho_{t-1}
\end{array}}\sum_{\Gamma_{t,T}}\sum_{\Gamma_{t+T,S}}\mathbb{P}\left(\Gamma_{t+T,S}|\varrho_{t+T-1}\right)\times\\
 & \mathbb{P}\left(\Gamma_{t,T}|\varrho_{t}\right)\mathrm{Tr}\left\{ \mathbf{\Psi}\left(\mathbf{\Psi}\left(x\mathbf{I},\Gamma_{t,T}\right),\Gamma_{t+T,S}\right)\right\} .
\end{align*}
Using~(\ref{eq:aux-3}) and the concavity of $\mathbf{\Psi}(\cdot,\Gamma_{t+T,S})$~\cite[Lemma 1e]{sinopoli2004kfi},
we have 
\begin{align*}
 & \xi_{t,T+S}(x)\\
\leq & \sup_{\varrho_{t+T-1}}\sum_{\Gamma_{t+T,S}}\mathbb{P}\left(\Gamma_{t+T,S}|\varrho_{t+T-1}\right)\times\\
\times & \mathrm{Tr}\left\{ \mathbf{\Psi}\left(\Xi(x),\Gamma_{t+T,S}\right)\right\} ,
\end{align*}
with 
\[
\Xi(x)=\sup_{\varrho_{t-1}}\sum_{\Gamma_{t,T}}\mathbb{P}\left(\Gamma_{t,T}\left|\varrho_{t-1}\right.\right)\mathbf{\Psi}\left(x\mathbf{I},\Gamma_{t,T}\right).
\]
Now, since $\mathbf{\Psi}(x\mathbf{I},\Gamma_{t,T})\geq0$, 
\begin{eqnarray*}
\Xi(x) & \leq & \sup_{\varrho_{t-1}}\sum_{\Gamma_{t,T}}\mathbb{P}\left(\Gamma_{t,T}\left|\varrho_{t-1}\right.\right)\mathrm{Tr}\left\{ \mathbf{\Psi}\left(x\mathbf{I},\Gamma_{t,T}\right)\right\} \mathbf{I}\\
 & = & \xi_{t,T}(x)\mathbf{I}.
\end{eqnarray*}
Then, using~(\ref{eq:aux-3}), it follows that 
\begin{eqnarray*}
\xi_{t,T+S}(x) & \leq & \sup_{\varrho_{t+T-1}}\sum_{\Gamma_{t+T,S}}\mathbb{P}\left(\Gamma_{t+T,S}|\varrho_{t+T-1}\right)\times\\
 & \times & \mathrm{Tr}\left\{ \mathbf{\Psi}\left(\xi_{t,T}(x)\mathbf{I},\Gamma_{t+T,S}\right)\right\} \\
 & = & \xi_{t+T,S}\left(\xi_{t,T}(x)\right)=\xi_{t+T,S}\circ\xi_{t,T}(x).
\end{eqnarray*}
\end{pf}
The reason why the map $\xi_{t,T}(x)$ is particularly important for
our analysis is because of the following result, which states a sufficient
condition for $G$ to be finite. 
\begin{lem}
\label{lem:EEC-bound-1} If there exists $T\in\mathbb{N}$ and $\bar{x}>0$
such that \textbf{$\max_{0\leq t<\tau}\limsup_{k\rightarrow\infty}\xi_{t,kT}(x)\leq\bar{x}<\infty$},
for all $x>0$, then 
\[
G<\infty.
\]
\end{lem}
\begin{pf}
Following the steps in~\cite[Lemma 30]{rohr2014kalman}, we can show
that 
\begin{equation}
\left\Vert \mathbb{E}\left(\mathbf{\Psi}\left(\mathbf{P}_{t},\Gamma_{t,T}\right)\right)\right\Vert \leq\xi_{t,T}(\|\mathbf{P}_{t}\|).\label{boaideia}
\end{equation}
Now, from~(\ref{boaideia}) Lemmas~\ref{lem:upper-bound} and~\ref{lem:xi}
\begin{align*}
 & \left\Vert \mathbb{E}\left(\mathbf{\Psi}\left(\mathbf{P}_{t},\Gamma_{t,kT+S}\right)\right)\right\Vert \\
\leq & \xi_{t,kT+S}(\|\mathbf{P}_{t}\|)\\
\leq & \xi_{t+kT,S}\circ\xi_{t,kT}(\|\mathbf{P}_{t}\|)\\
= & \sup_{\varrho_{s-1}}\sum_{\Gamma_{s,S}}\mathbb{P}\left(\Gamma_{s,S}\left|\varrho_{s-1}\right.\right)\mathrm{Tr}\left\{ \mathbf{\Psi}\left(\xi_{t,kT}(\|\mathbf{P}_{t}\|)\mathbf{I},\Gamma_{s,S}\right)\right\} \\
\leq & n\sup_{\varrho_{s-1}}\sum_{\Gamma_{s,S}}\mathbb{P}\left(\Gamma_{s,S}\left|\varrho_{s-1}\right.\right)\left\Vert \mathbf{\Psi}\left(\xi_{t,kT}(\|\mathbf{P}_{t}\|)\mathbf{I},\Gamma_{s,S}\right)\right\Vert \\
\leq & n\sup_{\varrho_{s-1}}\sum_{\Gamma_{s,S}}\mathbb{P}\left(\Gamma_{s,S}\left|\varrho_{s-1}\right.\right)\times\\
\times & \left(\xi_{t,kT}(\|\mathbf{P}_{t}\|)|\alpha_{1}|^{2S}c_{1}s^{2(\bar{J}-1)}+l_{S}\right)\\
= & n\xi_{t,kT}(\|\mathbf{P}_{t}\|)|\alpha_{1}|^{2S}c_{1}s^{2(\bar{J}-1)}+nl_{S}.
\end{align*}
Hence 
\begin{align*}
 & \max_{0\leq t<\tau}\limsup_{T\rightarrow\infty}\left\Vert \mathbb{E}\left(\mathbf{\Psi}\left(\mathbf{P}_{t},\Gamma_{t,T}\right)\right)\right\Vert \\
\leq & \sup_{0<S<T}n|\alpha_{1}|^{2S}c_{1}s^{2(\bar{J}-1)}\times\\
\times & \max_{0\leq t<\tau}\limsup_{k\rightarrow\infty}\xi_{t,kT}(\|\mathbf{P}_{t}\|)+nl_{S}\\
\leq & \sup_{0<S<T}n|\alpha_{1}|^{2S}c_{1}s^{2(\bar{J}-1)}\bar{x}+nl_{S}.
\end{align*}
Finally, since the above bound is independent of $t$ and$\mathbf{P}_{t}$,
\[
G\leq\sup_{0<S<T}n|\alpha_{1}|^{2S}c_{1}s^{2(\bar{J}-1)}\bar{x}+nl_{S}<\infty.
\]
\end{pf}

\subsubsection{Second step\label{sec:second_suff_cond}}

An alternative sufficient condition for the ANEEC $G$ to be bounded
is now presented. It will be shown in Section~\ref{subsec:equivalence}
that this condition is equivalent to the sufficient condition in Theorem~\ref{thm:main}. 
\begin{notation}
\label{not:41} For $k=1,\cdots,K$, let 
\[
{\cal N}_{k,Q}^{t,T}=\{\Gamma_{t,T}:\mathbf{O}_{k}(\Gamma_{t,T})\text{ does not have FCR}(Q)\},
\]
Notice that ${\cal N}_{k,0}^{t,T}={\cal N}_{k}^{t,T}$. 
\end{notation}
The following lemma presents a sufficient condition for the ANEEC
to be bounded. 
\begin{lem}
\label{lem:suff_cond} Under Assumption~\ref{assumption:1}, there
exists $Q\in\mathbb{N}_{0}$ such that, if 
\begin{equation}
|\alpha_{k}|^{2}\max_{0\leq t<\tau}\limsup_{T\rightarrow\infty}\mathbb{P}\left({\cal N}_{k,Q}^{t,T}\right)^{1/T}<1~\mathrm{for~all}~k=1,\cdots,K,\label{eq:assumpthm}
\end{equation}
then $G<\infty$. 
\end{lem}
\begin{pf}
The proof is divided into 5 steps.

1) In view of Lemma~\ref{lem:num_th}, there exists $Q$ such that
if $\mathbf{O}_{k}\left(\Gamma_{t,T}\right)$ has FCR$(Q)$, for $k=1,\cdots,K$,
then $\mathbf{O}\left(\Gamma_{t,T}\right)$ has FCR. Recall from Notation~\ref{not:41}
that ${\cal N}_{k,Q}^{t,T}$ is the set of sequences $\Gamma_{t,T}\in\mathcal{A}^{T}$
such that $\mathbf{O}_{k}(\Gamma_{t,T})$ does not have FCR with strength
$Q$. Define the set 
\begin{equation}
{\cal G}_{k,Q}^{t,T}\triangleq\begin{cases}
{\cal N}_{1,Q}^{t,T}, & k=1,\\
\bigcap\limits _{j=1}^{k-1}\overline{{\cal N}_{j,Q}^{t,T}}\cap{\cal N}_{k,Q}^{t,T}, & k=2,\cdots,K,
\end{cases}\label{eq:def_G}
\end{equation}
where $\overline{{\cal X}}$ denotes the complement of the set ${\cal X}$.
Notice that ${\cal G}_{k,Q}^{t,T}$ is the set of sequences $\Gamma_{t,T}\in\mathcal{A}^{T}$
such that $\mathbf{O}_{j}(\Gamma_{t,T})~\text{has~FCR}(Q)$ for $1\leq j\leq k-1$
and $\mathbf{O}_{k}(\Gamma_{t,T})~\text{does~not~have~FCR}(Q)$. Hence
\begin{equation}
{\cal G}_{k,Q}^{t,T}\subseteq{\cal N}_{k,Q}^{t,T}.\label{eq:P<P}
\end{equation}
Let ${\cal N}^{t,T}$ be the set of sequences $\Gamma_{t,T}\in\mathcal{A}^{T}$
such that $\mathbf{O}(\Gamma_{t,T})$ does not have FCR. From Lemma~\ref{lem:num_th},
we have 
\[
\mathcal{A}^{T}=\overline{{\cal N}^{t,T}}\cup\bigcup\limits _{k=1}^{K}{\cal G}_{k,Q}^{t,T},\,\forall t\in\mathbb{Z}.
\]
From Lemma~\ref{lem:upper-bound}, there exists $c_{1}>0$ such that
\begin{eqnarray*}
\Gamma_{t,T}\in\overline{{\cal N}^{t,T}} & \implies & \|\mathbf{\Psi}(x\mathbf{I},\Gamma_{t,T})\|\leq l_{T}\\
\Gamma_{t,T}\in{\cal G}_{k}^{t,T} & \implies & \|\mathbf{\Psi}(x\mathbf{I},\Gamma_{t,T})\|\leq|\alpha_{k}|^{2T}c_{1}T^{2(\bar{J}-1)}x+l_{T}.
\end{eqnarray*}

2) Recall that $\mathbf{\Psi}(x\mathbf{I},\Gamma_{t,T})$ is a symmetric,
positive-definite matrix with dimension $n$. Hence 
\[
\mathrm{Tr}\left(\mathbf{\Psi}(x\mathbf{I},\Gamma_{t,T})\right)\leq n\left\Vert \mathbf{\Psi}(x\mathbf{I},\Gamma_{t,T})\right\Vert .
\]
From~\eqref{eq:def_xi_3} and the above, we have 
\begin{align}
 & \xi_{t,T}(x)\nonumber \\
\leq & n\sup_{\varrho_{t-1}}\left\Vert \sum_{\Gamma_{t,T}}\mathbb{P}\left(\Gamma_{t,T}\left|\varrho_{t-1}\right.\right)\mathbf{\Psi}(x\mathbf{I},\Gamma_{t,T})\right\Vert \nonumber \\
\leq & n\sup_{\varrho_{t-1}}\sum_{\Gamma_{t,T}\in\overline{{\cal N}^{t,T}}}\mathbb{P}\left(\Gamma_{t,T}\left|\varrho_{t-1}\right.\right)\left\Vert \mathbf{\Psi}(x\mathbf{I},\Gamma_{t,T})\right\Vert +\nonumber \\
+ & n\sum_{k=1}^{K}\sup_{\varrho_{t-1}}\sum_{\Gamma_{t,T}\in{\cal G}_{k}^{t,T}}\mathbb{P}\left(\Gamma_{t,T}\left|\varrho_{t-1}\right.\right)\left\Vert \mathbf{\Psi}(x\mathbf{I},\Gamma_{t,T})\right\Vert .\label{eq:limxi}
\end{align}
Let 
\begin{align*}
A_{t,T} & =\sup_{\varrho_{t-1}}\sum_{\Gamma_{t,T}\in\overline{{\cal N}^{t,T}}}\mathbb{P}\left(\Gamma_{t,T}\left|\varrho_{t-1}\right.\right),\\
B_{t,T,k} & =\sup_{\varrho_{t-1}}\sum_{\Gamma_{t,T}\in{\cal G}_{k}^{t,T}}\mathbb{P}\left(\Gamma_{t,T}\left|\varrho_{t-1}\right.\right),\\
C_{t,T,k} & =\sup_{\varrho_{t-1}}\sum_{\Gamma_{t,T}\in{\cal N}_{k,Q}^{t,T}}\mathbb{P}\left(\Gamma_{t,T}\left|\varrho_{t-1}\right.\right).
\end{align*}
From~(\ref{eq:P<P}), $B_{t,T,k}\leq C_{t,T,k}$. Then, from~\eqref{eq:limxi},
Lemma~\ref{lem:upper-bound} and the above, we have 
\begin{align}
 & \xi_{t,T}(x)\nonumber \\
\leq & nl_{T}A_{t,T}+n\sum_{k=1}^{K}\left(|\alpha_{k}|^{2T}c_{1}T^{2(\bar{J}-1)}x+l_{T}\right)B_{t,T,k}\nonumber \\
\leq & nl_{T}A_{t,T}+n\sum_{k=1}^{K}\left(|\alpha_{k}|^{2T}c_{1}T^{2(\bar{J}-1)}x+l_{T}\right)C_{t,T,k}\nonumber \\
= & n\sum_{k=1}^{K}\left(|\alpha_{k}|^{2}C_{t,T,k}^{1/T}\right)^{T}c_{1}T^{2(\bar{J}-1)}x+\nonumber \\
+ & nl_{T}\left(A_{t,T}+\sum_{k=1}^{K}C_{t,T,k}\right)\nonumber \\
= & \beta_{t,T}x+\varphi_{t,T},\label{eq:aux-4}
\end{align}
with 
\begin{align*}
\beta_{t,T} & =nc_{1}T^{2(\bar{J}-1)}\sum_{k=1}^{K}\left(|\alpha_{k}|^{2}C_{t,T,k}^{1/T}\right)^{T},\\
\varphi_{t,T} & =nl_{T}\left(A_{t,T}+\sum_{k=1}^{K}C_{t,T,k}\right).
\end{align*}

3) Recall the definition of $\zeta$ from~(\ref{eq:independence}).
We have
\begin{align*}
C_{t,T,k} & =\sup_{\varrho_{t-1}}\sum_{\Gamma_{t,T}\in{\cal N}_{k,Q}^{t,T}}\mathbb{P}\left(\Gamma_{t,T}\left|\varrho_{t-1}\right.\right)\\
 & =\sup_{\varrho_{t-1}}\sum_{\substack{\Gamma_{t,T}\in{\cal N}_{k,Q}^{t,T}\\
\mathbb{P}\left(\Gamma_{t,T}\right)\neq0
}
}\frac{\mathbb{P}\left(\Gamma_{t,T}|\varrho_{t-1}\right)}{\mathbb{P}\left(\Gamma_{t,T}\right)}\mathbb{P}\left(\Gamma_{t,T}\right)\\
 & \leq\zeta\sum_{\substack{\Gamma_{t,T}\in{\cal N}_{k,Q}^{t,T}\\
\mathbb{P}\left(\Gamma_{t,T}\right)\neq0
}
}\mathbb{P}\left(\Gamma_{t,T}\right)\\
 & =\zeta\mathbb{P}\left({\cal N}_{k,Q}^{t,T}\right).
\end{align*}
Hence, 
\begin{equation}
\beta_{t,T}\leq\zeta nc_{1}T^{2(\bar{J}-1)}\sum_{k=1}^{K}\left(|\alpha_{k}|^{2}\mathbb{P}\left({\cal N}_{k,Q}^{t,T}\right)^{1/T}\right)^{T}.\label{eq:aux}
\end{equation}

4) From~(\ref{eq:assumpthm}), 
\begin{equation}
\max_{0\leq t<\tau}\limsup_{T\rightarrow\infty}|\alpha_{k}|^{2}\mathbb{P}\left({\cal N}_{k,Q}^{t,T}\right)^{1/T}<1.\label{eq:aux-6}
\end{equation}
In view of~(\ref{eq:aux-6}), there exists $\bar{T}\in\mathbb{N}$
such $\bar{\beta}\triangleq\max_{0\leq t<\tau}\beta_{t,\bar{T}}<1$.
We also have that, for all $0\leq t<\tau$, 
\begin{equation}
\varphi_{t,\bar{T}}\leq\bar{\varphi}\triangleq nl_{\bar{T}}\left(1+K\right).\label{eq:aux-5}
\end{equation}
Let $x_{k+1}=\bar{\beta}x_{k}+\bar{\varphi}$. We have that, for any
$x_{0}$, 
\[
\lim_{k\rightarrow\infty}x_{k}\leq\frac{\bar{\varphi}}{1-\bar{\beta}}.
\]
Also, from~(\ref{eq:aux-4}), $\xi_{t,k\bar{T}}(x_{0})\leq x_{k}$,
for all $0\leq t<\tau$. Hence 
\[
\max_{0\leq t<\tau}\limsup_{k\rightarrow\infty}\xi_{t,k\bar{T}}(x_{0})\leq\lim_{k\rightarrow\infty}x_{k}\leq\frac{\bar{\varphi}}{1-\bar{\beta}}<\infty,
\]
and the result follows from Lemma~\ref{lem:EEC-bound-1}. 
\end{pf}

\subsubsection{Third step\label{subsec:equivalence}}

The main goal of this section is to show that, for any $q\in\mathbb{N}_{0}$
and $k\in\{1,\cdots,K\}$, 
\[
\max_{0\leq t<\tau}\limsup_{T\rightarrow\infty}\mathbb{P}\left({\cal N}_{k,q}^{t,T}\right)^{1/T}=\max_{0\leq t<\tau}\limsup_{T\rightarrow\infty}\mathbb{P}\left({\cal N}_{k}^{t,T}\right)^{1/T}.
\]
Since in our analysis the value of $k$ is fixed, we remove it from
the notation. 
\begin{defn}
Let $\mu,\nu\in\mathcal{M}\left(\mathcal{E}\right)$. We say that
$\nu$ is absolutely continuous with respect to $\mu$ ($\nu\ll\mu$)
if $\nu(A)=0$ whenever $\mu(A)=0$. We say that $\mu$ and $\nu$
are mutually singular ($\mu\perp\nu$) if there exist $A\in\mathcal{B}$
such that $\mu(B)=0$ for all $A\supset B\in\mathcal{B}$ and $\nu\left(B\right)=0$
for all $\mathcal{E}\setminus A\supset B\in\mathcal{B}$.
\end{defn}
\begin{lem}
[Lebesgue's decomposition theorem] For $\mu,\nu\in\mathcal{M}\left(\mathcal{E}\right)$,
there is a unique decomposition $\mu=\mu_{\nu}+\mu_{\nu}^{\perp}$
with $\mu_{\nu}\ll\nu$ and $\mu_{\nu}^{\perp}\perp\nu$.
\end{lem}
\begin{lem}
\label{lem:perron-ext} Let $\mathcal{U}\subset\mathcal{M}\left(\mathcal{E}\right)$
be a closed subspace and $\kappa\in\mathcal{L}\left(\mathcal{U}\right)$.
If for every non-zero non-negative $\mu\in\mathcal{U}$ and $A\in\mathcal{F}\left(\mathcal{U}\right)$,
the following two conditions hold
\begin{enumerate}
\item $\kappa\mu(A)\geq0$,
\item there exists $N$ such that $\kappa\mu(A)>0$, for all $n\geq N$, 
\end{enumerate}
then, for every non-zero non-negative $\mu\in\mathcal{U}$, 
\[
\lim_{n\rightarrow\infty}\left\Vert \kappa^{n}\mu\right\Vert ^{1/n}=\rho(T).
\]

\end{lem}
\begin{pf}
We split the proof in steps:

1) Using the transfinite recursion theorem and Lebesgue's decomposition
theorem, we can construct a family $\mathcal{Q}\left(\mathcal{U}\right)\subset\mathcal{P}\left(\mathcal{E}\right)$
of mutually singular probability measures such that, for all $\mu\in\mathcal{M}\left(\mathcal{E}\right)$,
\[
\mu\perp\nu\text{ for all }\nu\in\mathcal{U}\Leftrightarrow\mu\perp\nu\text{ for all }\nu\in\mathcal{Q}\left(\mathcal{U}\right).
\]
We then have that, for every $\mu\in\mathcal{U}$,
\[
\mu=\bigcup_{\nu\in\mathcal{Q}\left(\mathcal{U}\right)}\mu_{\nu}.
\]
Hence, $\mathcal{U}$ is isomorphic to the $l_{1}$-sum of the spaces
$L_{1}\left(\mathcal{E},\mathcal{B},\nu\right)$, for all $\nu\in\mathcal{Q}\left(\mathcal{U}\right)$,
i.e., 
\[
\mathcal{U}\cong\left(\bigoplus_{\nu\in\mathcal{Q}\left(\mathcal{U}\right)}L_{1}\left(\mathcal{E},\mathcal{B},\nu\right)\right)_{l_{1}}.
\]
It then follows that the dual space $\mathcal{U}^{\ast}$ of $\mathcal{U}$
is 
\begin{equation}
\mathcal{U}^{\ast}\cong\left(\bigoplus_{\nu\in\mathcal{Q}\left(\mathcal{U}\right)}L_{\infty}\left(\mathcal{E},\mathcal{B},\nu\right)\right)_{l_{\infty}}.\label{eq:dual}
\end{equation}

2) It follows from~(\ref{eq:def-F}) that 
\[
\mathcal{F}\left(\mathcal{U}\right)=\mathcal{F}\left(\mathcal{Q}\left(\mathcal{U}\right)\right).
\]
Hence, conditions~1 and~2 imply that, for every $\nu\in\mathcal{Q}\left(\mathcal{U}\right)$
and non-zero positive $f\in L_{\infty}\left(\mathcal{E},\mathcal{B},\nu\right)$,
$\int fd(\kappa\mu)_{\nu}\geq0$ and there exists $N$ such that $\int fd(\kappa^{n}\mu)_{\nu}>0$,
for all $n>N$. Then, it follows from~(\ref{eq:dual}) that these
two conditions also follow for any non-zero non-negative $f\in\mathcal{U}^{\ast}$.
Also, it is straightforward to verify that, when equipped with the
total variation norm, and the natural partial order, $\mathcal{M}(\mathcal{E})$
is a Banach lattice. Then, in the terminology of~\cite{sawashima1964spectral},
the first condition means that $\kappa$ is a positive operator and
the second one means that it is non-support.

3) Let
\[
R\left(\lambda,\kappa\right)=\left(\lambda I-\kappa\right)^{-1},\text{ for all }\,\lambda\in\mathbb{C},
\]
be the resolvent of $\kappa$. It follows from the conclusion of~2)
and~\cite[Corollary on p. 61]{sawashima1964spectral} that $\rho(\kappa)$
is a pole of $R\left(\lambda,\kappa\right)$ with multiplicity $1$.
Hence, from~\cite[Th. 2.3 (e)]{marek1970frobenius}, the operator
\[
P\triangleq\lim_{n\rightarrow\infty}\frac{\kappa^{n}}{\rho^{n}(\kappa)},
\]
is well defined (i.e., the limit converges in the operator norm) and
there exists $N$ such that, for all $n>N$, 
\[
\int fd\left(P^{n}\mu\right)\neq0,
\]
for every $\nu\in\mathcal{Q}\left(\mathcal{U}\right)$ and every non-zero
positive $f\in L_{\infty}\left(\mathcal{E},\mathcal{B},\nu\right)$.
The above implies that we must have
\[
P\mu\neq0.
\]
We then have
\begin{align*}
\lim_{n\rightarrow\infty}\left\Vert \kappa^{n}\mu\right\Vert ^{1/n} & =\lim_{n\rightarrow\infty}\left\Vert \frac{\kappa^{n}}{\rho^{n}(\kappa)}\mu\right\Vert ^{1/n}\rho(\kappa)\\
 & =\lim_{n\rightarrow\infty}\left\Vert P\mu\right\Vert ^{1/n}\rho(\kappa)\\
 & =\rho(\kappa).
\end{align*}
\end{pf}
\begin{notation}
\label{nota:4} Fix $k$ and recall Notation~\ref{nota:2}. Put $\Gamma^{(1)}=\Gamma$,
and for each $l\in\mathbb{N}$, consider the following iterations
\begin{align*}
n_{l} & =\min\left\{ n:\left(\Gamma^{(l)}(1),\cdots,\Gamma^{(l)}(nM)\right)\notin\mathcal{N}_{k}^{nM}\right\} ,\\
\Gamma^{(l+1)} & =\left(\Gamma^{(l)}(n_{l}M+1),\cdots,\Gamma^{(l)}(|\Gamma^{(l)}|)\right),
\end{align*}
where $|\Gamma|$ denotes the length of the sequence $\Gamma$. The
iterations are stopped at $l=L$, where $L$ is such that $\Gamma^{(L+1)}\in\mathcal{N}_{k}^{\left|\Gamma^{(L+1)}\right|}$.
Define the maps 
\begin{align*}
\eta\left(\Gamma\right) & =L,\\
\tau\left(\Gamma\right) & =\Gamma^{(L+1)},
\end{align*}
For each $t,n,q\in\mathbb{N}_{0}$, let $\lambda_{t,n,q}:\mathcal{I}\rightarrow\mathcal{M}\left(\mathcal{E}\right)$
be defined by
\begin{multline*}
\lambda_{t,n,q}\left(i\right)(A)=\mathbb{P}\left(\varrho_{t+nM}\in A,\psi\left(\tau\left(\Gamma_{t,nM}\right)\right)=\mathcal{K}_{i},\right.\\
\left.\eta\left(\Gamma_{t,nM}\right)=q\right)
\end{multline*}
\end{notation}
\begin{rem}
\label{rem:clarification}The above notation can be interpreted as
follows. Suppose that we start processing blocks of $M$ contiguous
measurements starting from time $t$. Using these measurements we
build an observability matrix. Whenever this matrix has FCR, we leave
it aside and restart building a new matrix with the next block. After
processing $n$ blocks, $\eta\left(\Gamma_{t,nM}\right)$ denotes
the number of FCR matrices so accumulated and $\tau\left(\Gamma_{t,nM}\right)$
denotes the sequence of blocks remaining after removing those used
to build FCR matrices. Also, $\psi\left(\tau\left(\Gamma_{t,nM}\right)\right)$
denotes the kernel of the observability matrix build with these remaining
blocks. Then, $\lambda_{t,n,q}\left(i\right)(A)$ denotes the probability
that, after $n$ blocks, we accumulated $q$ FCR matrices, the kernel
induced by the remaining blocks is $\mathcal{K}_{i}$ and the final
Markov state $\varrho_{t+nM}$ belongs to the set $A$.
\end{rem}
\begin{lem}
\label{lem:sc2}For any $Q\in\mathbb{N}_{0}$ and $t\in\mathbb{Z}$
\[
\limsup_{T\rightarrow\infty}\mathbb{P}\left({\cal N}_{Q}^{t,T}\right)^{1/T}\leq\max_{0\leq i<I}\rho\left(\breve{\varsigma}_{t}(i,i)\right)^{1/M},
\]
with equality holding when $Q=0$. 
\end{lem}
\begin{pf}
Recall Notations~\ref{nota:2} and~\ref{nota:4}. We split the proof
in steps:

1) In view of the order given to the kernels $\mathcal{K}_{i}$, it
follows that, for $i\in\mathcal{I}\setminus\{0\}$,
\begin{equation}
\lambda_{t,n,q}\left(i\right)=\sum_{j=0}^{i}\varsigma_{t}\left(i,j\right)\lambda_{t,n-1,q}\left(j\right),\label{eq:L1}
\end{equation}
and for $i=0$,
\begin{equation}
\lambda_{t,n,q}\left(0\right)=\sum_{j=0}^{I}\varsigma_{t}\left(I,j\right)\lambda_{t,n-1,q-1}\left(j\right)+\varsigma_{t}\left(0,0\right)\lambda_{t,n-1,q}\left(0\right).\label{eq:L2}
\end{equation}

Let $\mathbf{l}_{t,n}\in\mathcal{M}^{QI}\left(\mathcal{E}\right)$
be defined by $\mathbf{l}_{t,n}=\left[\mathbf{l}_{t,n,0}^{\top},\cdots,\mathbf{l}_{t,n,Q}^{\top}\right]^{\top}$
with $\mathbf{l}_{t,n,q}=\left[\lambda_{t,n,q}(0),\cdots,L_{t,n,q}(I-1)\right]^{\top}$.
Then, from~(\ref{eq:L1})-(\ref{eq:L2}),
\begin{equation}
\mathbf{l}_{t,n}=\mathbf{B}_{t}\mathbf{l}_{t,n-1},\label{eq:prob-trans}
\end{equation}
where 
\[
\mathbf{B}_{t}=\left[\underbrace{\begin{array}{cccc}
\mathbf{D}_{t} & \mathbf{0} & \cdots & \mathbf{0}\\
\mathbf{M}_{t} & \mathbf{D}_{t} & \ddots & \vdots\\
\mathbf{0} & \ddots & \ddots & \mathbf{0}\\
\mathbf{0} & \mathbf{0} & \mathbf{M}_{t} & \mathbf{D}_{t}
\end{array}}_{q+1\text{-times}}\right],
\]
with 
\begin{align*}
\mathbf{D}_{t} & =\left[\begin{array}{ccc}
\varsigma_{t}\left(0,0\right) & \mathbf{0} & \mathbf{0}\\
\vdots & \ddots & \mathbf{0}\\
\varsigma_{t}\left(I-1,1\right) & \cdots & \varsigma_{t}\left(I-1,I-1\right)
\end{array}\right],\\
\mathbf{M}_{t} & =\left[\begin{array}{ccc}
\varsigma_{t}\left(I,1\right) & \cdots & \varsigma_{t}\left(I,I-1\right)\\
\mathbf{0} & \cdots & \mathbf{0}
\end{array}\right].
\end{align*}
(Notice that, since $M$ is multiple of $\tau$, then, probability
transitions in~(\ref{eq:prob-trans}) are independent of $n$.) We
then have 
\begin{equation}
\mathbf{l}_{t,n}=\mathbf{B}_{t}^{n}\mathbf{l}_{t,0},\label{eq:sc0}
\end{equation}

2) Let
\[
\left\Vert \mathbf{l}_{t,n}\right\Vert \triangleq\sum_{q=0}^{Q}\sum_{i=0}^{I-1}\left\Vert \lambda_{t,n,q}\left(i\right)\right\Vert .
\]
We then have 
\[
\mathbb{P}\left(\eta\left(\Gamma_{t,nM}\right)\leq q\right)=\left\Vert \mathbf{l}_{t,n}\right\Vert =\left\Vert \mathbf{B}_{t}^{n}\mathbf{l}_{t,0}\right\Vert ,
\]
We also have 
\begin{align*}
\eta\left(\Gamma_{t,nM}\right) & >q\Rightarrow\Gamma_{t,nM}\notin{\cal N}_{q}^{t,nM},\\
\eta\left(\Gamma_{t,nM}\right) & >0\Leftrightarrow\Gamma_{t,nM}\notin{\cal N}^{t,nM}.
\end{align*}
Hence, using~(\ref{eq:sc0}), for any $n\in\mathbb{N}_{0}$, 
\begin{equation}
\mathbb{P}\left({\cal N}_{q}^{t,nM}\right)\leq\left\Vert \mathbf{l}_{t,n}\right\Vert =\left\Vert \mathbf{B}_{t}^{n}\mathbf{l}_{t,0}\right\Vert ,\label{eq:sc1}
\end{equation}
with equality holding when $q=0$.

3) The matrix representation of the operator $\mathbf{B}_{t}:\mathcal{M}^{QI}\left(\mathcal{E}\right)\rightarrow\mathcal{M}^{QI}\left(\mathcal{E}\right)$
is lower triangular, and so are its diagonal entries $\mathbf{D}_{t}$.
We the have that the spectrum $\sigma\left(\mathbf{B}_{t}\right)$
of $\mathbf{B}_{t}$ satisfies 
\[
\sigma\left(\mathbf{B}_{t}\right)=\bigcup_{i=0}^{I-1}\sigma\left(\varsigma_{t}(i,i)\right).
\]
Hence, the spectral radius $\rho\left(\mathbf{B}_{t}\right)$ of $\mathbf{B}_{t}$
is
\[
\rho\left(\mathbf{B}_{t}\right)=\max_{0\leq i<I}\rho\left(\varsigma_{t}(i,i)\right).
\]

4) From Gelfand's formula~\cite[eq. (5.2-5)]{taylor1958introduction},
\begin{align}
\limsup_{n\rightarrow\infty}\left\Vert \mathbf{B}_{t}^{n}\mathbf{l}_{t,0}\right\Vert ^{1/n} & \leq\limsup_{n\rightarrow\infty}\left\Vert \mathbf{B}_{t}^{n}\right\Vert ^{1/n}\left\Vert \mathbf{l}_{t,0}\right\Vert ^{1/n}\nonumber \\
 & =\rho\left(\mathbf{B}_{t}\right)\nonumber \\
 & =\max_{0\leq i<I}\rho\left(\varsigma_{t}(i,i)\right)\nonumber \\
 & =\max_{0\leq i<I}\rho\left(\breve{\varsigma}_{t}(i,i)\right),\label{eq:sc5}
\end{align}
where the last equality follows from~(\ref{eq:restriction}). Also,
\begin{align}
\limsup_{n\rightarrow\infty}\left\Vert \mathbf{B}_{t}^{n}\mathbf{l}_{t,0}\right\Vert ^{1/n} & \geq\max_{0\leq i<I}\limsup_{n\rightarrow\infty}\left\Vert \varsigma_{t}^{n-i}(i,i)\mu_{t,i}\right\Vert ^{1/n}\nonumber \\
 & =\max_{0\leq i<I}\limsup_{n\rightarrow\infty}\left\Vert \varsigma_{t}^{n}(i,i)\mu_{t,i}\right\Vert ^{1/n},\label{eq:sc7}
\end{align}
with 
\[
\mu_{t,i}=\varsigma_{t}(j,j-1)\cdots\varsigma_{t}(1,0)\mu_{0}.
\]
Let
\[
\left(\mathbf{C}(\varrho),\mathbf{R}(\varrho)\right)\triangleq h(\varrho),
\]
Also, for each $i\in\mathcal{I}$ and $m=1,\cdots,M$, let $\mathcal{D}_{i}=\left(D_{i,m}\in\mathcal{B}:m=1,\cdots,M\right)$,
where
\[
D_{i,m}=\left\{ \varrho\in\mathcal{E}:\ker\left(\mathbf{C}(\varrho)\mathbf{A}^{m}\right)\supset\mathcal{K}_{i}\right\} .
\]
We have that
\[
\varsigma_{t}(i,i)=\prod_{m=1}^{M}\chi_{D_{i,m}}\kappa_{t+m}(i,i).
\]
Let $\mathcal{V}=\mathcal{U}\left(\varsigma_{t}(i,i)\right)$. It
follows from~(\ref{eq:non-support}) that $\breve{\varsigma}_{t}(i,i):\mathcal{V}\rightarrow\mathcal{V}$
(recall~\ref{eq:breve-opp}) satisfies the conditions of Lemma~\ref{lem:perron-ext}.
Hence,
\begin{align*}
\limsup_{n\rightarrow\infty}\left\Vert \varsigma_{t}^{n}(i,i)\mu_{t,i}\right\Vert ^{1/n}\geq & \limsup_{n\rightarrow\infty}\left\Vert \breve{\varsigma}_{t}^{n}(i,i)\chi_{\mathcal{V}}\mu_{t,i}\right\Vert ^{1/n}\\
= & \rho\left(\breve{\varsigma}_{t}(i,i)\right).
\end{align*}
We then have, from~(\ref{eq:sc7}),
\begin{equation}
\limsup_{n\rightarrow\infty}\left\Vert \mathbf{B}_{t}^{n}\mathbf{l}_{t,0}\right\Vert ^{1/n}\geq\max_{0\leq i<I}\rho\left(\breve{\varsigma}_{t}(i,i)\right).\label{eq:sc6}
\end{equation}
Then, from~(\ref{eq:sc5}) and~(\ref{eq:sc6}),
\begin{equation}
\limsup_{n\rightarrow\infty}\left\Vert \mathbf{B}_{t}^{n}\mathbf{l}_{t,0}\right\Vert ^{1/n}=\max_{0\leq i<I}\rho\left(\breve{\varsigma}_{t}(i,i)\right).\label{eq:sc4}
\end{equation}

5) Let $n(T)=\max\{n\in\mathbb{N}:nM\leq T\}$. We have 
\begin{align}
\limsup_{T\rightarrow\infty}\mathbb{P}\left({\cal N}_{q}^{t,T}\right)^{1/T} & \leq\limsup_{T\rightarrow\infty}\mathbb{P}\left({\cal N}_{q}^{t,n(T)M}\right)^{\frac{1}{(n(T)+1)M}}\nonumber \\
 & =\limsup_{n\rightarrow\infty}\mathbb{P}\left({\cal N}_{q}^{t,nM}\right)^{\frac{1}{(n+1)M}}\nonumber \\
 & =\limsup_{n\rightarrow\infty}\mathbb{P}\left({\cal N}_{q}^{t,nM}\right)^{\frac{1}{nM}}.\label{eq:sc2}
\end{align}
But also 
\begin{align}
\limsup_{T\rightarrow\infty}\mathbb{P}\left({\cal N}_{q}^{t,T}\right)^{1/T} & \geq\limsup_{T\rightarrow\infty}\mathbb{P}\left({\cal N}_{q}^{t,(n(T)+1)M}\right)^{\frac{1}{nM}}\nonumber \\
 & =\limsup_{n\rightarrow\infty}\mathbb{P}\left({\cal N}_{q}^{t,nM}\right)^{\frac{1}{nM}}.\label{eq:sc3}
\end{align}
Then, from~(\ref{eq:sc1}),~(\ref{eq:sc4}),~(\ref{eq:sc2}) and~(\ref{eq:sc3})
\begin{align*}
\limsup_{T\rightarrow\infty}\mathbb{P}\left({\cal N}_{q}^{t,T}\right)^{1/T} & =\limsup_{n\rightarrow\infty}\mathbb{P}\left({\cal N}_{q}^{t,nM}\right)^{\frac{1}{nM}}\\
 & \leq\limsup_{n\rightarrow\infty}\left\Vert \mathbf{B}_{t}^{n}\mathbf{l}_{t,0}\right\Vert ^{\frac{1}{nM}}\\
 & =\max_{0\leq i<I}\rho\left(\breve{\varsigma}_{t}(i,i)\right)^{1/M},
\end{align*}
with equality when $q=0$. 
\end{pf}
\begin{lem}
\label{lem:eqP}For any $q\in\mathbb{N}_{0}$ and $t\in\mathbb{Z}$,
\[
\limsup_{T\rightarrow\infty}\mathbb{P}\left({\cal N}_{q}^{t,T}\right)^{1/T}=\limsup_{T\rightarrow\infty}\mathbb{P}\left({\cal N}^{t,T}\right)^{1/T}.
\]
\end{lem}
\begin{pf}
We have ${\cal N}^{t,T}\subseteq{\cal N}_{q}^{t,T}$. Hence, using
Lemma~\ref{lem:sc2} we obtain, 
\begin{align*}
\limsup_{T\rightarrow\infty}\mathbb{P}\left({\cal N}^{t,T}\right)^{1/T} & \leq\limsup_{T\rightarrow\infty}\mathbb{P}\left({\cal N}_{q}^{t,T}\right)^{1/T}\\
 & \leq\rho_{t}^{1/M}\\
 & =\limsup_{T\rightarrow\infty}\mathbb{P}\left({\cal N}^{t,T}\right)^{1/T},
\end{align*}
and the result follows. 
\end{pf}
We are now ready to prove the sufficient condition in Theorem~\ref{thm:main}. 
\begin{pf}
{[}of the sufficient condition in Theorem~\ref{thm:main}{]} The
sufficient condition in Theorem~\ref{thm:main}, i.e.,~\eqref{eq:only_if_part},
follows immediately from Lemmas~\ref{lem:suff_cond} and~\ref{lem:eqP}. 
\end{pf}

\section{Conclusion\label{sec:conclusion}}

We stated a necessary and sufficient condition for stability of a
Kalman filter under general assumptions on the linear system and its
random measurement equation. We also studied how to numerically compute
this condition for a given system. Furthermore, we used our result
to assess the stability in a networked setting involving sensor scheduling
and packet dropouts. This shows how our stability condition is a rather
general one that could be applied in a widely range of applications,
including those found in networked control settings.

\appendix

\section{Proof of Proposition~\ref{prop:deg} and~Corollary~\ref{cor:non-deg}\label{sec:proof_lemmas}}
\begin{pf}
{[}of Proposition~\ref{prop:deg}{]} It follows immediately from
Lemma~\ref{lem:sc2} by making $q=0$. 
\end{pf}
\begin{pf}
{[}of Proposition~\ref{prop:spec-rad}{]} It follows from~(\ref{eq:non-support})
that $\breve{\varsigma}_{t}(i,i):\mathcal{U}\left(\varsigma_{t}(i,i)\right)\rightarrow\mathcal{U}\left(\varsigma_{t}(i,i)\right)$
satisfies the conditions of Lemma~\ref{lem:perron-ext}. The result
then follows immediately from that lemma.
\end{pf}
\begin{pf}
{[}of Corollary~\ref{cor:non-deg}{]}  Let $\left(\mathcal{E},\mathcal{B}\right)$
be defined as in Remark~\ref{rem:HMM}. Let $\Gamma_{N}^{\ast}\in\mathcal{\mathcal{A}}^{N}$
be defined by
\[
\Gamma_{N}^{\ast}(n)=\left(\mathbf{C}^{(\alpha)},\mathbf{R}^{(\beta)}\right),\text{ for all }n=1,\cdots,N,
\]
and $\Gamma_{\infty}^{\ast}\in\mathcal{\mathcal{A}}^{\mathbb{N}}$
be the extension of this sequence to $\mathcal{\mathcal{A}}^{\mathbb{N}}$.
We have that $\mathbb{K}=\left\{ \mathcal{K}_{1}\right\} $, i.e.,
it has only one element given by
\[
\mathcal{K}_{1}=\mathrm{ker}\left(O\left(\Gamma_{N}^{\ast}\right)\right).
\]
Let $\mathcal{D}=\left(D_{m}\in\mathcal{B}:m=1,\cdots,M\right)$ with
\[
D_{m}=\left\{ \left(\mathbf{C}^{(\alpha)},\mathbf{R}^{(\beta)}\right)\right\} .
\]
Let also
\[
\eta_{t}=\prod_{m=1}^{M}\chi_{D_{m}}\kappa_{t+m},
\]
We then have
\begin{align*}
\mathcal{U}\left(\eta_{t}\right) & =\left\{ a\delta_{\Gamma_{\infty}^{\ast}}:a\in\mathbb{R}\right\} ,\\
\mathcal{F}\left(\mathcal{U}\left(\eta_{t}\right)\right) & =\left\{ \Gamma_{\infty}^{\ast}\right\} .
\end{align*}
We then have $\breve{\eta}_{t}:\mathcal{U}\left(\eta_{t}\right)\rightarrow\mathcal{U}\left(\eta_{t}\right)$
is defined by
\[
\breve{\eta}_{t}\left(\delta_{\Gamma_{\infty}^{\ast}}\right)=p\delta_{\Gamma_{\infty}^{\ast}}
\]
where
\[
p=\prod_{t=0}^{\tau-1}\mathbb{P}\left(\mathbf{C}_{t}=\mathbf{C}^{(\alpha)}|\mathbf{C}_{s}=\mathbf{C}^{(\alpha)},s<t\right).
\]
Hence,~(\ref{eq:non-support}) holds provided $p>0$. 

It is easy to see that, for any $\lambda\in\mathbb{C}$, the map $\left(\lambda I-\eta_{t}\right)$
has an inverse unless $\lambda=0$ or $\lambda=p$. Hence, $\sigma\left(\eta_{t}\right)=\left\{ 0,p\right\} $.
It is also immediate that $\sigma\left(\breve{\eta}_{t}\right)=\left\{ p\right\} $.
Hence
\[
\rho\left(\eta_{t}\right)=\rho\left(\breve{\eta}_{t}\right)=p,
\]
and~(\ref{eq:restriction}) holds. We can them use~(\ref{eq:alg_claim})
to obtain the result, after noticing that
\[
\rho\left(\varsigma_{t}(i,i)\right)=\rho\left(\eta_{t}\right)^{M/\tau}=p^{M/\tau}.
\]
\end{pf}
\bibliographystyle{unsrt}
\bibliography{refs}

\end{document}